\documentstyle[11pt]{article}
\font\cmss=cmss10 
\font\cmsss=cmss10 at 7pt

\def\Z{\relax\ifmmode\mathchoice {\hbox{\cmss Z\kern-.4em Z}}
      {\hbox{\cmss Z\kern-.4em Z}} {\lower.9pt\hbox{\cmsss Z\kern-.4em Z}} 
      {\lower1.2pt\hbox{\cmsss Z\kern-.4em Z}}\else{\cmss Z\kern-.4emZ}\fi}
\def\p{\partial} 
\def\D{{\cal D}}

\def\e{\epsilon} 
\def\g{\gamma}
\def\n{\nabla} 
\def\d{\delta} 
\def\s{\sigma}

\def\b{\beta} 
\def\a{\alpha} 
\def\l{\lambda}
\def\f{\varphi} 

\def\da{{\dot\alpha}} 
\def\db{{\dot\beta}}
\def\dg{{\dot\gamma}}

\def\wt{\widetilde}
\def\A{{\cal A}}
\def\M{{\cal M}}

\def\op{\oplus}
\def\om{\ominus} 
\def\ot{\otimes}
\def\inbar{\,\vrule height1.5ex width.4pt depth0pt}
\def\C{\relax\,\hbox{$\inbar\kern-.3em{\rm C}$}}
\def\II{\relax{\rm I\kern-.17em I}}
\def\I1{\relax{\rm 1\kern-.28em l}}
\def\IP{\relax{\rm I\kern-.18em P}}
\def\R{\relax{\rm I\kern-.18em R}}
\def\half{{\textstyle{1\over 2}}}
\def\3h{{\textstyle{3\over 2}}} 
\def\N#1{$N\!=\!#1$ }
\def\der#1{{\partial \over \partial #1}}

\def\be{\begin{equation}}
\def\r#1{(\ref{#1})} 
\def\la#1{\label{#1}}
\def\c#1{\cite{#1}}
\def\ee{\end{equation}}
\def\arr{\begin{array}{rll}}
\def\ea{\end{array}}
\def\bea{\begin{eqnarray}}
\def\eea{\end{eqnarray}}
\def\bealph#1
  {\setcounter{equation}{0}\renewcommand\theequation{#1\alph{equation}}\bea}
\def\eealph#1
  {\eea\setcounter{equation}{#1}\renewcommand\theequation{\arabic{equation}}}
\def\Mi{Minkowski space}
\def\sM{superspace}
\def\ss{superspace}
\def\sd{self-dual}
\def\asd{antiself-dual}
\def\sdy{self-duality}
\def\asdy{antiself-duality}

\def\ssdy{super self-duality}

\def\sp{super-Poincar\'e algebra}
\def\sym{super Yang-Mills} 
\def\ym{Yang-Mills}
\begin{document}
\baselineskip=15pt plus 1pt minus 1pt
{\small
\rightline{hep-th/9704036}
}
\vskip 2.0 true cm
\thispagestyle{empty}
{\Large
\centerline{{Supersymmetric Lorentz-Covariant Hyperspaces}} 
\centerline{{and self-duality equations in dimensions greater than (4$|$4)}}
}
\vskip 1.5 true cm

\centerline{{\Large{C.~Devchand\,}}\footnote{e-mail: devchand@ictp.trieste.it}}
\vspace{.1in}
\centerline{ International Centre for Theoretical Physics,
34100 Trieste, Italy}
\centerline{ Max-Planck-Institut f\"ur Gravitationsphysik,
14473 Potsdam, Germany}
\vspace{.3in}
\centerline{and}
\vspace{.3in}
\centerline{{\Large{Jean~Nuyts\,}}{\footnote{e-mail: Jean.Nuyts@umh.ac.be}}}
\vspace{.1in}
\centerline{ Physique Th\'eorique et Math\'ematique, 
Universit\'e de Mons-Hainaut}
\centerline{20 Place du Parc, 7000 Mons, Belgium}
\vskip 2.0 true cm

\begin{abstract}
{We generalise the notions of supersymmetry and superspace by allowing 
generators and coordinates transforming according to more general Lorentz
representations than the spinorial and vectorial ones of standard lore.
This yields novel SO(3,1)-covariant superspaces, which we call hyperspaces,
having dimensionality greater than $(4|4)$ of traditional super-Minkowski 
space. As an application, we consider gauge fields on complexifications of 
these superspaces; and extending the concept of self-duality, we obtain classes 
of completely solvable equations analogous to the four-dimensional self-duality
equations. 
}
\end{abstract}
\newpage

\section{Introduction}
 
\vskip 12pt \noindent{\bf 1.1\ }
{\it Supersymmetry} and {\it self--duality} have both yielded very 
fruitful geometric concepts for recent developments in field theory, 
string physics and differential geometry. It seems that generalisations
of both these ideas provide a broader framework for possible 
further applications, both mathematical and physical. The purpose of this 
paper is to describe certain generalisations of these notions.
Specifically, we consider generalised superspaces, 
which we in general call {\em hyperspaces},
coordinated by some finite subset from the set of general Lorentz tensors 
$\{ Y^{ \a_1 \dots \a_{2a}\da_1 \dots \da_{2\dot a}}\} $ with 
$a=0,\half,\dots$ and $\dot a=0,\half,\dots\ $, in standard 
two-spinor notation.
These tensors are separately symmetrical in their $2a$ undotted
and $2\dot a $ dotted indices and transform according to the
$(a,\dot a)$ representation of the Lorentz group.
This is a method of parametrising 
spaces of arbitrary dimensionality in a manifestly four-dimensional
Lorentz covariant fashion. 
We introduce gauge fields on such hyperspaces
and, on complexifications of these spaces, extend the notion of 
{\it self--duality} by requiring certain irreducible components
of the curvature tensor to vanish, just as the familiar \sdy\ condition
is tantamount to the vanishing of the $(0,1)$ component 
of the field strength tensor.

\vskip 12pt \noindent{\bf  1.2\ }
Our construction of hyperspaces is modeled on standard
superspace. The latter is constructed as a coset space: the super-Poincar\'e
group over the Lorentz group. This follows the description of \Mi\ as the
coset of the Poincar\'e group by the Lorentz group. Now, factoring out the
Lorentz group from the super-Poincar\'e group yields a space of dimension
$(4|4)$ with four odd (fermionic) coordinates $Y^\a, Y^\da$ 
transforming according to
the spinorial $(\half,0),(0,\half)$ representations of the Lorentz group
in addition to the four even (bosonic) vectorial $(\half,\half )$ 
coordinates $Y^{\a\da}$ of standard \Mi.
A representation of the super Poincar\'e algebra 
is given by vector fields on \sM.
The super translation vector fields $( X_{\a},X_{\db},X_{\a\db} )$,
built with the same odd ($(\half,0)$ and $(0,\half)$) and even
($(\half,\half)$) representations as the corresponding coordinates,
realise the superalgebra
\be
\left\{  X_{\a} , X_{\db}\right\}  =\  2i\  X_{\a\db}\  ,
\label{st1} 
\ee
with all other supercommutators 
(i.e. commutators between two even tensors and between one even
and one odd tensor and anticommutators between two odd tensors)
equal to zero. These vector fields together with the elements of the Lorentz 
algebra realise the super-Poincar\'e algebra. In the standard coordinate
basis, they have the following non-zero 
supercommutation relations with the \sM\ coordinates,
\be
\arr
\left[  X_{\a\da}, Y^{\b\db} \right]    =  \d^\b_\a  \d^\db_\da\  ,\quad
\left\{  X_{\a} , Y^{\b } \right\}   &=&   \d^\b_\a\  ,\quad
\left\{  X_{\da}, Y^\db  \right\}   =  \d^\db_\da \\[5pt]
\left[  X_{\a} , Y^{\b\db} \right]    = i\  \d^\b_\a Y^\db \  ,&\quad&
\left[  X_{\da}, Y^{\b\db}\right]   =  i  \d^\db_\da Y^{\b}\  \  ,
\la{st2}
\ea
\ee
in virtue of which, the $X$'s can be realised in terms of a holonomic
basis of partial derivatives with respect to the $Y$'s thus:
\be  
X_{\a\db} = \der{Y^{\a\db}} ,\quad 
X_{\a}= \der{Y^\a} + iY^\db \der{Y^{\a\db}} ,\quad 
X_{\db}= \der{Y^\db} + iY^\a \der{Y^{\a\db}}\  .\la{st3}
\ee

Now, in the standard case \c{hls} of the super Poincar\'e algebra, one insists
that the commutator of an 
odd element ($(\half,0)$ or $(0,\half)$)
with the even element ($(\half,\half)$) vanishes. 
In particular, no $(1,{1\over 2})$ element $X_{\a\b\db}$ is allowed to appear
in the fashion:
\be
\left[  X_{\a} , X_{\b\db} \right]   = X_{\a\b\db} + \dots 
\ \ \ .
\ee
This restriction, however, can be lifted and
we may indeed think of the $(\half,0)$ and $(0,\half)$ 
elements as generating
successively higher-spin elements of a generalised superalgebra. So, 
for instance, the further action of the $(\half,0)$ element on the
$(1,\half)$ can yield a $({3\over 2},\half)$ element,
\be
\left\{   X_{\a} , X_{\b\g\db} \right\}   = X_{\a\b\g\db} + \dots 
\ee
and so on. The new elements $X_{\a\b\db},X_{\a\b\g\db},\dots$
may be realised as vector fields on a generalised superspace, with 
coordinates $Y^{\a\b\da} , Y^{\a\b\g\da},\dots$ beyond the traditional
$(\half,0), (0,\half)$ and $(\half,\half)$ representations. 
These coordinates are interpreted 
as coordinates of extra dimensions (in possibly a Kaluza--Klein sense)
of a higher-dimensional superspace. 
The Lorentz invariance, however, remains that of four-dimensional space. 
This is therefore a way of going to a higher-dimensional space whilst
maintaining the four-dimensional Lorentz structure fixed to the coordinate
system. For instance, the
simple bosonic extension of four-dimensional space with coordinates
$( Y^{\a\da} , Y^{\a\b\g\da} )$ has dimension $4+8 = 12$, or one with
coordinates $(Y^{\alpha\dot\alpha},Y^{\alpha\beta\dot\alpha\dot\beta})$
has dimension 4+9 =13.

We thus consider generalised superalgebras $\A\ $, with elements $X$
taken as a finite subset of the set of general Lorentz tensors of the form
$\left\{   X_{ \a_1\dots \a_{2a}\da_1\dots \da_{2\dot a}}\right\}  $ with 
$a=0,\half \dots$ and $\dot a =0,\half \dots\ $; 
the same representations possibly appearing more than once.
These tensors are, like the corresponding coordinates $Y$, separately 
symmetrical in their $2a$ undotted and $2\dot a$ dotted indices and 
transform according to the $(a,\dot a)$ representation of the Lorentz group. 
Thus including elements of higher Lorentz spin, represented by vector
fields on correspondingly generalised superspaces, we generalise the idea 
of standard \ss. More precisely, our superalgebras $\A\ $ can be 
represented by vector fields on 
hyperspaces $\M\ $ with coordinates in one-to-one 
correspondence with the elements of $\A\ $ modulo perhaps the Lorentz 
generators, analogously to the realisation of \sM\ with coordinates 
$(Y^{\a\da}, Y^\a, Y^\da)$ in one-to-one correspondence with the 
elements of the \sp\ modulo the Lorentz algebra. 

We shall assume that all Lorentz tensors 
$\left\{  T(a,\dot a)= T_{ \a_1\dots \a_{2a}\da_1\dots \da_{2\dot a}}\right\} $ 
take values in a $\Z_2$-graded (super) vector space.
The degree or parity of a tensor with respect to the $\Z_2$-grading,
$d(T)$, is defined by $2(a+\dot a)$ mod 2 and
$T$ will be called bosonic (and taken to be grassmann-even)
if $d(T)=0$ and fermionic (and taken to be grassmann-odd) if $d(T)=1$.
The supercommutator or graded bracket between two tensors is defined by 
\be
\left[ A(a,\dot a),B(b,\dot b)\right] = 
A(a,\dot a)B(b,\dot b) - (-1)^{d(A)d(B)} B(b,\dot b)A(a,\dot a)\ .
\label{eta}
\ee
It is automatically graded skew-symmetric,
\be
\left[ A(a,\dot a),B(b,\dot b)\right] 
  = - (-1)^{d(A)d(B)} \left[ B(b,\dot b),A(a,\dot a)\right]\ , 
\ee
and satisfies the super Jacobi identity
\bea 
\left[ A(a,\dot a),\left[ B(b,\dot b),C(c,\dot c)\right] \right]  
&+& (-1)^{d(A)(d(B)+d(C))} \left[ B(b,\dot b),
\left[ C(c,\dot c),A(a,\dot a)\right] \right] 
\nonumber \\ 
&+& (-1)^{d(C)(d(A)+d(B))} \left[ C(c,\dot c),
\left[ A(a,\dot a),B(b,\dot b)\right] \right] 
=0\ .
\la{superjacobi}\eea
When both $A$ and $B$ are known to be odd, we shall follow the custom  
of denoting this bracket by the anticommutator $\{A, B\}$.
Although our framework is rather
more general than that in which the spin--statistics theorem is valid,
we assume that objects with an odd (resp. even) number of (dotted plus 
undotted) indices are Grassmann--odd or fermionic (resp. Grassmann-even 
or bosonic). We note that this is not necessarily the case.
For instance Lie- (rather than super-) algebra extensions of the 
d-dimensional Poincar\'e algebra acting on spaces with Grassmann--even 
spinorial coordinates have recently been classified \c{ac}.

The (Lorentz invariant) structure constants of the algebra $\A\ $ 
encode the nonholonomic nature (torsion)
of the vector fields $X$ on $\M$. By solving the 
super-Jacobi identities for the $X$'s,
we shall construct some explicit examples of higher-spin algebras and 
corresponding superspaces. We shall not {\it a priori} insist 
on Poincar\'e symmetry, but consider the most general associative 
superalgebras generated by these Lorentz tensors.
The action of the $X$'s on the $Y$'s is given by supercommutation 
relations generalising \r{st2} and consistent choices of coordinate 
bases afford determination by solving the super-Jacobi identities 
involving supercommutations of $X$'s  with  $Y$'s.

Our consideration is more general than that of \c{hls} in that we
allow the appearance of vector fields of spin greater than one
and although our construction is Lorentz-covariant, we moreover
do not, a priori, demand four-dimensional Poincar\'e invariance.    
The classification of \c{hls}, which was restricted to extensions of
Poincar\'e symmetry in four-dimensions, was generalised in \c{nahm} 
to higher dimensions, where  the supersymmetrisations of 
{\it higher-dimensional} Poincar\'e (and de-Sitter) symmetry were 
considered. Our generalisation to higher dimensions, on the other 
hand, maintains manifest {\it four-dimensional} Lorentz covariance. 
In that we allow the existence of generators of spin greater than one, 
our consideration is close in spirit to that of Fradkin and Vasiliev 
\c{fv}, who were concerned with realising 
higher-spin 
superalgebras $\A\ $ 
on fields in four dimensional de Sitter space. They considered the 
higher-spin generators as giving rise to higher-spin fields, whose 
consistent dynamics 
in the curved de Sitter space described by the the spin 2 field,
however, required $\A\ $ to be infinite dimensional, 
with an action on a chain of fields having spins all the way up to infinity. 
We however interpret the higher-spin generators of $\A\ $ as momenta in 
extra dimensions coordinated by higher-spin coordinates. 
We do not make any a priori field theoretical or dynamical requirements.
In particular, we realise our algebra in flat space. The super Jacobi
identities therefore
afford any number of finite dimensional solutions and 
the maximal spin can be chosen at will. 

\vskip 12pt \noindent{\bf 1.3\ }
In this paper we do not pursue the interesting possibilities for lagrangian
field theories on our hyperspaces $\M$, nor do we attempt to
describe higher-spin dynamics using the algebras $\A$. There are many such
exciting possibilities for future work, extending, for instance, the early
considerations of Fierz \c{f} on higher-spin dynamics, or the more recent 
investigations of Fradkin and Vasiliev \c{fv,v} on the realisation of 
higher-spin superalgebras
$\A\ $ on interacting fields including gravity. We restrict 
ourselves here to one simple field-theoretical application:
We consider gauge fields on the hyperspaces $\M$.
Since the vector fields $X$ act as superderivations on functions of $Y$,  
they can be gauge covariantised by adding
a  gauge potential 
$A$ transforming according to the same representation of the
Lorentz group as $X$. Commutators of gauge-covariantised vector fields,
i.e. of the $\A$-covariant derivatives,
then yield curvatures which decompose into irreducible representations 
of the Lorentz group. 
Without pursuing the question of lagrangian field theories for such 
generalised gauge fields,  we presently investigate the possibility
of generalising the very fruitful notion of {\it self--duality}
to hyperspaces $\M$. 
This yields interesting classes of solvable gauge-invariant systems 
in superspaces of basically arbitrary dimensionality. 
 
Euclidean space self-duality equations have played a central role 
in the search for classical solutions to gauge theories in 
virtue of transforming the second--order field equations into simpler 
first--order ones. Originally introduced in four dimensional spaces,
the idea of considering algebraic curvature
constraints as a means to solving the second--order Yang-Mills equations
was extended in a natural way to euclidean spaces of higher dimensions in 
\cite{CDFN}, where systems of first-order equations for the gauge 
potential were constructed, which imply the second-order Yang-Mills
equations and which are invariant under some subgroup $H$ of the 
d-dimensional rotation group SO(d). This 
generalisation of \sdy\ concerned the construction of a fourth-rank 
$H$-invariant tensor $T_{\mu\nu\rho\s}$ which could be used instead of
the four-dimensional SO(4) invariant tensor $\e_{\mu\nu\rho\s}$. Then, 
the eigenvalue equations for the tensor $T$, namely, 
\be  
T_{\mu\nu\rho\s} F^{\rho\s} = \l F_{\mu\nu}\ \ \,
\ee
generalise four-dimensional \sdy\ 
in that they are algebraic curvature constraints 
which imply the \ym\ equations in virtue of the Bianchi identities. 
Projections to distinct eigenspaces of $T$, with eigenvalues $\{ \l \}$,
correspond to generalisations of self- and anti-self-dual parts of the
curvature. This construction therefore
generalises the role of the four-dimensional Hodge-duality operator 
as an endomorphism of the space of two-forms
with self- and anti-dual eigenspaces.
In the present paper, however,  we generalise four dimensional \sdy\  
in another direction.

In two-spinor notation the commutator of two covariant derivatives
manifestly displays the irreducible representation 
content of the gauge field:
\be 
\left[ \D_{\a\da} ~,~ \D_{\b\db}\right]  =\  \e_{\da\db} F_{\a\b} 
                                  + \e_{\a\b} F_{\da\db}\  .
\label{curv}
\ee
The imposition of \sdy\ (resp. \asdy) is just a statement of 
the vanishing of the (0,1) component $F_{\da\db}$ 
(resp. the (1,0) component $F_{\a\b}$).
Equivalently, one can say that the selfdual curvature
contains only the (1,0) Lorentz representation, i.e. 
\be 
\left[ \D_{\a\da} ~,~ \D_{\b\db}\right]  =\  \e_{\da\db} F_{\a\b}
\quad\Leftrightarrow\quad
F_{\da\db} =  0\ , 
\label{sd3}
\ee
and analogously for the antiselfdual case.
As these equations show, the use of two-spinor notation is not only a
very convenient way of manifestly displaying the irreducible parts of
the field-strength tensor, but this decomposition is also revealed to 
be equivalent to the decomposition in eigenstates of the Hodge--duality 
operator. This equivalence is central to many of the wonderful mathematical
properties of the \sdy\ equations. 
So instead of using the above-mentioned
$T$-tensor construction of \c{CDFN}, we could equally generalise 
the alternative notion that {\it \sdy} corresponds to the absence of certain 
irreducible representations in the decomposition of 
generalised curvature tensors on $\M$.
The imposition of such `coherent' curvature constraints on $\M$ 
thus maintains the usual four-dimensional 
rotation group SO(4) as a basic symmetry of the equations.  
It is this generalisation of \sdy\ which we pursue in this paper; this
generalisation being more immediately applicable to superspaces with 
both odd and even parts than the T-tensor construction. 

Self-duality equations have recently drawn renewed attention as  
unifying systems for lower-dimensional integrable equations \c{w2}
and it has been suggested that the twistor transform could be 
the `mother' of lower dimensional transforms which render the latter 
completely integrable (like the inverse scattering transform). Many
coherent curvature  constraints on $\M$ also arise as 
integrability conditions for linear systems and the advantage
of manifest four-dimensional covariance is that a generalised twistor--type
transformation is easily constructed. We shall discuss some 
interesting classes of linear systems, generalising not only those
for the \sdy\ and anti-\sdy\ equations, but also the linear system
for the conventional supercurvature constraints of superspace
\ym\ theories. The latter, for the \N3 extension,  we recall,
are equivalent to the full \sym\ equations \c{hhls}. Our discussion
of linear systems for coherent curvature constraints generalises  
Ward's approach to completely solvable curvature constraints
in dimensions greater than four \c{w}.   

\vskip 12pt \noindent{\bf  1.4\ }
The plan of this paper is as follows. In section 2 we introduce
coordinates and $\A$-covariant translation vector fields on $\M$, 
which satisfy, for consistency, the super-Jacobi identities
discussed in section 3. An explicit novel example of a hyperspace $\M$,
containing coordinates and covariant derivatives
with spins up to ${3\over 2}$
is presented in section 4. The introduction of gauge fields on such
superspaces is discussed in section 5. Curvature constraints generalising
\sdy\ and integrability conditions for them are discussed in section 6
and explicit examples are given. We deal mainly with
`N=1' superspaces $\M$, with coordinates of any given Lorentz type
appearing only once. Some details of the extension to examples with certain 
Lorentz representations appearing multiply is given in appendix \ref{mult}.
The analogues of \sdy\ discussed in section 6 are soluble in the sense that
the familiar \sdy\ equations are: in virtue of a twistor transform to
freely specifiable holomorphic data.
 
We use two-spinor language with dotted and
undotted indices raised and lowered by the skew-symmetric symplectic
invariant tensors $\e_{\a\b},\e_{\da\db},\e^{\a\b},\e^{\da\db}$,
with $\e_{12}=1=\e^{21}$.  The generators of Lorentz transformations
satisfy
\be\arr
 \left [ M_{\alpha\beta},
       M_{\gamma\delta}\right ]&=&
    \epsilon_{\alpha\gamma}M_{\beta\delta}
   +\epsilon_{\alpha\delta}M_{\beta\gamma}
   +\epsilon_{\beta\gamma}M_{\alpha\delta}
   +\epsilon_{\beta\delta}M_{\alpha\gamma} 
\\[6pt]
\left [M^{\dot{\alpha}\dot{\beta}},
       M^{\dot{\gamma}\dot{\delta}}\right ] & = &
    \epsilon^{\dot{\alpha}\dot{\gamma}}M^{\dot{\beta}\dot{\delta}}
   +\epsilon^{\dot{\alpha}\dot{\delta}}M^{\dot{\beta}\dot{\gamma}}
   +\epsilon^{\dot{\beta}\dot{\gamma}}M^{\dot{\alpha}\dot{\delta}}
   +\epsilon^{\dot{\beta}\dot{\delta}}M^{\dot{\alpha}\dot{\gamma}}  
\\[6pt]
\left [M_{\alpha\beta},
       M^{\dot{\gamma}\dot{\delta}}\right ]&=&0 
\ea
\label{lorentz2} 
\ee
with $ M_{\a\b} $ and $ M^{\da\db} $ acting respectively on {\em undotted} 
and {\em dotted} indices. 
The 2-spinor notation is particularly suited to the description 
of half-integer spin representations. In
fact, it brings to light the fact that when the algebra of the
Lorentz group is extended by allowing combinations of 
generators with complex coefficients, rather than real ones, 
the algebra can be split formally into two commuting
$SU(2)$'s, i.e. that the complex
extension of the Lorentz group, $SO(4,\C)$, is locally isomorphic to
 $SL(2,\C)\times SL(2,\C)$. 
For physical applications, care needs to be taken in imposing appropriate
hermiticity conditions at the end so that the theory 
transforms according to the appropriate real form of $SO(4,\C)$. 
In particular, in the Lorentzian case, when the four-dimensional 
$Y^{\a\da}$-subspace, $\M_4$, has (3,1) signature, the real form is
the simple group $SL(2,\C)$ and dotted and undotted indices are 
related by complex conjugation. This relation no longer holds
if the Lorentz group is taken to be either of the other possible
cases $SO(4) = SU(2)\times SU(2)$ (for $\M_4$ having euclidean (4,0) 
signature) or $SL(2,\R)\times SL(2,\R)$ (corresponding to an $\M_4$ 
with a kleinian (2,2) signature). 
We shall, however, deal mainly with the complex extension.
In particular, complexification lifts the Minkowski space
conjugation between dotted and undotted spinor representations and allows
the imposition of constraints like the above $ F_{\da\db} =\ 0  $,
leaving the choice of hermiticity conditions to be decided later
according to the physical application being considered.

We shall use the multi-index notation $[A],[B],[\dot A]$ and $[\dot B]$ 
to denote sets of, respectively, $2a,2b,2\dot a$ and $2\dot b$ symmetrized
 indices ($a, b, \dot a$ and $\dot b$ being integers or half-integers),
\be\arr
[A]=\alpha_1\alpha_2\ldots\alpha_{2a}\ \ ,&\quad&
[B]=\beta_1 \beta_2\ldots\beta_{2b}\ \  ,
 \\[5pt]
[\dot{A}]=\dot{\alpha_1}\dot{\alpha_2}
     \ldots\dot\alpha_{2\dot{a}}\ \ ,&\quad&
[\dot B] = \dot{\beta_1}\dot{\beta_2}\ldots \dot\beta_{2\dot b}
\ \  .\label{sets}
\ea\ee
Similarly $[A_p]$ (resp. $[\dot A_p]$)) will denote the set $[A]$ 
(resp. $[\dot A]$) with the index $\alpha_p$ (resp. $\da_p$) missing.
Using the $\epsilon$'s we also define multi-index epsilon tensors
\be
\epsilon_{[\a_s\b_s]}=\epsilon_{\alpha_1\beta_1}
                  \epsilon_{\alpha_2\beta_2}
                  \ldots
                  \epsilon_{\alpha_s\beta_s}  
\ \ ,\quad 
\epsilon^{[\da_{\dot s}\db_{\dot s}]}=
         \epsilon^{\dot\alpha_1\dot\beta_1}
         \epsilon^{\dot\alpha_2\dot\beta_2}
         \ldots
         \epsilon^{\dot\alpha_{\dot s}\dot\beta_{\dot s}}\ \ ,
\label{genep} 
\ee
where $ \epsilon^{[\a_0\b_0]}=\epsilon_{[\a_0\b_0]} 
= \epsilon^{[\da_0\db_0]}=\epsilon_{[\da_0\db_0]}=1$,
and their inverses
\be
\epsilon^{[\b_s\a_s]}=
       \epsilon^{\beta_{1}\alpha_{1}}
       \epsilon^{\beta_{2}\alpha_{2}}
       \ldots
       \epsilon^{\beta_{s}\alpha_{s}}
\ \ ,\quad 
\epsilon_{[\db_{\dot s}\da_{\dot s}]}=
         \epsilon_{\dot\beta_1\dot\alpha_1}
         \epsilon_{\dot\beta_2\dot\alpha_2}
         \ldots
         \epsilon_{\dot\beta_{\dot s}\dot\alpha_{\dot s}} \ \  ,
\label{epinv}
\ee
which satisfy $\epsilon^{[\b_s\a_s]} \epsilon_{[\a_s\b_s]}=  2^s$.
Moreover, we shall denote by $S[A]$, the
symmetrisation operator which symmetrically sums over
all the $(2a)!$ permutations of the indices in $[A]$.

Using this notation, an irreducible tensor 
$T(a,\dot a) \equiv T_{[A]}^{ [\dot A]}$, symmetric in its $2a$ undotted 
and  $2\dot a$ dotted indices transforms under (\ref{lorentz2}) as
\bea
\left [M_{\beta\gamma},T_{[A]}^{[\dot A]}\right ] & = &
         \sum_{p=1}^{2a} \epsilon_{\beta\alpha_p}
              T_{[A_p]\gamma}^{[\dot A]}        
        +\sum_{p=1}^{2a} \epsilon_{\gamma\alpha_p}
              T_{[A_p]\beta}^{[\dot A]}   
\nonumber \\
\left [M^{\dot{\beta}\dot{\gamma}},T_{[A]}^{[\dot A]}\right ]&=&
         \sum_{p=1}^{2\dot{a}} \epsilon^{\dot\beta{\dot\alpha_p}}
              T_{[A]}^{[\dot A_p]\dot\gamma}        
        +\sum_{p=1}^{2\dot{a}} \epsilon^{\dot\gamma{\dot\alpha_p}}
              T_{[A]}^{[\dot A_p]\dot\beta}         
\label{Ttrans} 
\eea
and corresponds to an irreducible $ (a,\dot a)$ representation of the 
Lorentz group having dimension  $(2a+1)(2\dot a+1)$. 
When using the multi-indices, we shall write, for visual clarity,
the undotted ones lowered and the dotted ones raised. Spinor indices can
of course always be raised or lowered Lorentz covariantly using the
epsilon tensors.  
The `spin' content of a tensor $T(a,\dot a)$
(i.e. its behaviour under `space rotation', the diagonal $su(2)$ 
algebra of (\ref{lorentz2})), is given by the decomposition
\be
a\otimes \dot a=(a+\dot a)\oplus (a+\dot a-1)
               \oplus \ldots\oplus \mid a-\dot a\mid
\ \ .
\label{spincontent}
\ee

\section{The higher-spin 
superalgebra $\A\ $ and the hyperspace $\M$}

\subsection{Hypersymmetries}

We consider a set of Lorentz tensors $\{X(a,\dot a)\}$, transforming according 
to (\ref{Ttrans}).  
Incorporating the nonholonomy \r{st1} of the \sp, 
we postulate, as defining relations for hypersymmetry algebras $\A$, 
the most general Lorentz-covariant super-commutation relations 
having a right-hand-side linear in the $X$'s:
\bea
&& \left[ X_{[A]}^{[\dot{A}]}~,~ X_{[B]}^{[\dot{B}]} \right]
\label{comXX} \\
  &=& \sum_{s=0}^{\min (2a,2b)}\sum_{\dot s=0}^{\min (2\dot a,2\dot b)}
    t(a,\dot a\,;b,\dot b\,;a+b-s,\dot a+\dot b-\dot s)
     S[A]S[\dot A]S[B]S[\dot B]
     \epsilon_{[\a_s\b_s]}\epsilon^{[\da_{\dot s}\db_{\dot s}]}
     X_{[C(s)]}^{[\dot C(\dot s)]}\nonumber
\eea 
where $t(a,\dot a\,;b,\dot b\,;c,\dot c)$ are structure constants
(the torsion, or more precisely anholonomy, coefficients) 
depending on six half-integers and we 
choose the convention that
$t(a,\dot a\,;b,\dot b\,;c,\dot c)=0$ 
when $c,\dot c$ are outside the range of the summation.
Here we have introduced 
the multi-indices (for $0\leq s \leq \min(2a,2b)$ and 
$0\leq \dot s \leq \min(2\dot a,2\dot b)$)
\be
\arr
[C(s)]&=&
          \alpha_{s+1}
          \alpha_{s+2}
          \ldots
          \alpha_{2a}
          \beta_{s+1}
          \beta_{s+2}
          \ldots
          \beta_{2b}  
  \\[5pt]
\left [\dot C(\dot s)\right ]&=&
               \dot\alpha_{\dot s+1}
               \dot\alpha_{\dot s+2}
               \ldots
               \dot\alpha_{2\dot a}
               \dot\beta_{\dot s+1}
               \dot\beta_{\dot s+2}
               \ldots
               \dot\beta_{2\dot b}\ \   
\label{Csets}
\ea
\ee
and we use the fact that the tensor product of irreducible Lorentz 
representations,
$(a,\dot a) \ot (b,\dot b) = (a\otimes b,\dot a\otimes \dot b)$,
decomposes according to the Clebsch-Gordan rules,
\be\arr    a\otimes b&=&(a+b)\oplus (a+b-1)
               \oplus \ldots\oplus \mid a-b\mid
\\[5pt]
\dot a\otimes \dot b&=&(\dot a+\dot b)\oplus (\dot a+\dot b-1)
               \oplus \ldots\oplus \mid \dot a-\dot b\mid\ ,
\label{tensorproduct}\ea\ee
with the Clebsch-Gordan coefficients for these spinor representations
being representable by the multi-index $\e$'s.  
The right-hand side of \r{comXX} clearly needs to have the symmetry 
properties of the left under the interchange of the indices 
$[A],[\dot A]$ with $[B],[\dot B]$. 
Taking into account the antisymmetry of the
$\epsilon$ factors, this leads to the following restrictions
on the $t$ parameters
\be
\arr
&& t(a,\dot a\,;b,\dot b\,;a+b-s,\dot a+\dot b-\dot s)
   \\[5pt]
&&=(-1)^{4(a+\dot a)(b+\dot b)+s+\dot s+1} 
t(b,\dot b\,;a,\dot a\,;a+b-s,\dot a+\dot b-\dot s)  
\label{tsym1}
\ea
\ee
We obtain, in particular, that
\be 
t(a,\dot a\,;a,\dot a\,;a-s,\dot a-\dot s)=0\ \  
\left\{\arr
&{\rm{if}}\ 2(a+\dot a)\ {\rm{is\ even\ and}}\ s+\dot s\
{\rm{is\ even}}           \\
&{\rm{if}}\ 2(a+\dot a)\ {\rm{is\ odd\ and}}\ \ s+\dot s\ 
{\rm{is\ odd}}  
\ea \right.
\label{tsym2}
\ee
Apart from the appropriate symmetry properties, associativity requires
the satisfaction of the relevant super 
Jacobi identities, which will be given in the next section. 
The operators $X$ then form a $Z_2$-graded superalgebra $\A$, with 
even (resp. odd) elements having even (resp. odd) $2(a+\dot a)$.  

We note that the natural identifications  
$X_{\a\b} =  M_{\a\b},\quad X^{\da\db} = M^{\da\db} $, with the Lorentz
generators may be made, though these are by no means necessary requirements.  
Further, elements  $X(a,\dot a)$ transforming according to any specific 
representation $(a, \dot a)$ could, in principle, occur multiply. 

\subsection{Hyperspaces} 

The simplest realisations of algebras ${\cal A}$ are as infinitesimal
translation vector fields on generalisations of standard superspace.
To this end, 
we enlarge \Mi\ to {\em hyperspace} $\M$, with coordinates
$Y(a,\dot a)$ corresponding to the algebra elements $X(a,\dot a)$
and also transforming according to \r{Ttrans}. 
For any given finite set  $\left\{  (a, \dot a)\right\} $, we interpret the 
(correspondingly even or odd) coordinates $\left\{  Y(a,\dot a)\right\} $ as coordinates 
of $(2a+1)(2\dot a+1)$-dimensional (even or odd) subspaces of $\M$.
Standard \sM, therefore, with coordinates $Y(\half,\half)$, 
$Y(\half,0)$ and $Y(0,\half)$, of subspaces of respectively
$4$ bosonic and $2+2$ fermionic dimension, has total dimension $(4|4)$;
and the super-Poincar\'e algebra has a manifestly covariant action on it.
On the other hand, \N4 extended \sM, with four copies of the odd
subspaces, has dimension $(4|16)$ and a covariant action of the \N4 
extended \sp. For the simplicity of our exposition, we relegate discussion
of analogous spaces $\M\ $, with certain representations appearing multiply,
to appendix \ref{mult}. 

As is usual for coordinates, we assume that they supercommute amongst 
themselves, i.e.
\be
\left [Y_{[A]}^{ [\dot A]},Y_{[B]}^{[\dot B]}\right ]=0\ . 
\label{comYY}
\ee

We now proceed to specify the action of the superalgebras $\A\ $ on 
superspaces $\M\ $ with coordinates $Y$. The vector fields $X\in \A\ $ 
clearly need to act as superderivations on functions of $Y$. To fulfil,
in particular, that the $X$'s map functions of $Y$ to functions of $Y$, 
we postulate, as the simplest possibility, that the action of an $X$
on a $Y$ is a linear combination of the $Y$'s, i.e. the hypersymmetry 
transforms the coordinates at most linearly among themselves. 
The $X$'s together with the $Y$'s therefore combine to form an
enlarged superalgebra,
with additional supercommutation relations  
\bea
&&\left[ X_{[A]}^{[\dot A]} ~,~ Y_{[B]}^{[\dot B]}\right ] 
\nonumber \\
&=&\sum_{s=0}^{\min (2a,2b)}\sum_{\dot s=0}^{\min (2\dot a,2\dot b)}
  u(a,\dot a\,;b,\dot b\,;a+b-s,\dot a+\dot b-\dot s)
S[A]S[\dot A]S[B]S[\dot B]\epsilon_{[\a_s\b_s]}
             \epsilon^{[\da_{\dot s}\db_{\dot s}]}
     Y_{[C(s)]}^{ [\dot C(\dot s)]}    
\nonumber \\
&&\quad      +\ c(a,\dot a) \delta_{ab} \delta^{\dot a\dot b}
   S[A]S[\dot A]\epsilon_{[\a_{2a}\b_{2a}]}
                \epsilon^{[\da_{2\dot a}\db_{2\dot a}]} \ .
\label{comXY}
\eea 
Additional structure constants have been labeled 
$u(a,\dot a\,;b,\dot b\,;c,\dot c)$ and essential central 
parameters $c(a,\dot a)$ have been introduced.
If any of the latter are non zero, they can always be renormalised 
to 1 by multiplying the $X$'s and/or the $Y$'s by an appropriate factor. 
Super-Jacobi identities yield quadratic consistency relations among
the structure constants 
$t(a,\dot a\,;b,\dot b\,;c,\dot c), u(a,\dot a\,;b,\dot b\,;c,\dot c)$
and $c(a,\dot a)$. These are discussed in the next section.

Given any particular set of $X$'s generating an algebra $\A\ $ 
with relations \r{comXX}, the span of $Y$'s satisfying \r{comYY},
\r{comXY} and the relevant Jacobi identities, can be thought of
as the set of coordinates of a $Z_2$-graded superspace $\M$.
For usual \Mi, the algebra $\A\ $ is generated by the Lorentz generators
together with the operators 
$X(\half,\half)$, which are simply realised as partial derivatives  
$\p / \p Y^{\a\da}$.
They commute among themselves (the $t$ parameters are zero) and their 
commutators with the coordinates involve only the central term 
$c(\half,\half)$, with the $u$ parameters being zero. 
Now just as \Mi\ can be thought of as the coset space
(Poincar\'e group)/(Lorentz group), we can consider 
the span of $Y$'s to be basically the coordinates of the supergroup 
corresponding to the algebra $\A$. 

Given a coordinate basis $\left\{  Y\right\} $, we clearly have a holonomic 
(supercommuting) basis of infinitesimal translation vector fields 
constructed from the partial derivatives 
\be
 \p_{[A]}^{[\dot A]} =S[A] S[\dot A] \frac{\p}{ \p Y^{[A]}_{[\dot A]}}
\label{partial}
\ee
satisfying \r{comXY} with all $t$'s and $u$'s set to zero
and all $c$'s put to unity
\be
\left[ \p_{[A]}^{[\dot A]} ~,~ Y_{[B]}^{[\dot B]}\right ] 
= \delta_{ab} \delta^{\dot a\dot b}
   S[A]S[\dot A]\epsilon_{[\a_{2a}\b_{2a}]}
                \epsilon^{[\da_{2\dot a}\db_{2\dot a}]}\ .
\label{compY}
\ee 
The relations \r{comXY} may be realised by $\A$-covariant or 
{\em hypercovariant} derivatives, using the notation \r{Csets},
\bea
X^{[\dot A]}_{[A]} &=& c(a,\dot a)\  \p^{[\dot A]}_{[A]} 
\nonumber \\
&& +\  \sum_{b=0} \sum_{\dot b=0} 
       \sum_{s=0}^{\min (2a,2b)} 
       \sum_{\dot s=0}^{\min (2\dot a,2\dot b)}
u(a,\dot a\,;b,\dot b\,;a+b-s,\dot a+\dot b-\dot s)
\nonumber \\  
&&\hskip 4.6 true cm
S[A] S[\dot A]
\epsilon_{[\a_s\b_s]}
\epsilon^{[\da_{\dot s}\db_{\dot s}]}
     Y_{[C(s)]}^{ [\dot C(\dot s)]} 
\partial^{[B]}_{[\dot B]}
\label{Xrepresent}
\eea
This coordinate realisation yields \r{comXY} straight away. However, 
requiring that these $X$'s satisfy \r{comXX} implies
quadratic relationships between
the $t$'s, the $u$'s and the $c$'s; relationships which, in the abstract
setting, arise from the super Jacobi identities among the $X$'s and $Y$'s. 
These relations do not necessarily have unique solution and particular
solutions correspond to particular choices of coordinate bases:
the standard non-chiral and chiral bases for \sM\ being the simplest
example (see section 4).

The algebra \r{comXX}, \r{comYY}, \r{comXY}  
is formally invariant under the following $\Z_2$ `chiral' transformations  
\be\arr
{\mbox{dotted upper index}} 
              &\leftrightarrow & {\mbox{undotted lower index}} \\[5pt] 
{\mbox{dotted lower index}} 
              &\leftrightarrow & {\mbox{undotted upper index}} \\[5pt]
t(a,\dot a\,;b,\dot b\,;c,\dot c)
              &\leftrightarrow &t(\dot a,a\,;\dot b,b\,;\dot c,c)\\[5pt]  
u(a,\dot a\,;b,\dot b\,;c,\dot c)
              &\leftrightarrow &u(\dot a,a\,;\dot b,b\,;\dot c,c)\\[5pt]  
c(a,\dot a)
              &\leftrightarrow &c(\dot a,a)\ \ \ .
\ea
\la{tildesym}
\ee
In particular, the number of independent structure constants and central
parameters $t$, $u$, and $c$ can be roughly halved
by imposing the {\it{maximal non-chirality}} condition of
$\Z_2$ chiral symmetry, i.e.
\be\arr
t(a,\dot a\,;b,\dot b\,;c,\dot c)&=&t(\dot a,a\,;\dot b,b\,;\dot c,c)
\\[5pt]  
u(a,\dot a\,;b,\dot b\,;c,\dot c)&=&u(\dot a,a\,;\dot b,b\,;\dot c,c)
\\[5pt]  
c(a,\dot a)&=&c(\dot a,a)\ \ \ .
\ea
\label{adasymmetry}
\ee

We note that the choice of the set of $Y$'s for a given $\A$ 
(or more generally of the $X$'s and the $Y$'s) is possibly not unique,
since non-linear transformations amongst the generators, 
preserving their tensorial nature
as well as the linearity of the right-hand sides, can be envisaged. 
We shall not pursue details of such equivalences; rather, we 
obtain the consistency conditions defining {\it all} algebras
having a given number of $X$'s and $Y$'s. 
Linear transformations among the generators, on the other hand, 
are ruled out by Lorentz covariance 
in the `$N=1$' cases of at most one element for each Lorentz behaviour.

It is straightforward to extend the algebra to the case where the 
multiplicity of certain Lorentz representations is greater than 
one. Then the structure constants $t$ and $u$ depend on three extra
multiplicity indices and the central terms $c$ on two extra
indices (i.e. they are matrices in the multiplicity space). In these 
extended cases the elements are defined up to linear transformations 
amongst the elements transforming similarly under Lorentz 
transformations. Such transformations can, moreover, be used to 
diagonalise the central terms. Details are given in appendix \ref{mult}.

\section{The Super Jacobi Identities}
We now derive the conditions implied by the super-Jacobi identities in
order that the vector fields $X$ and coordinates $Y$ form a 
super Lie algebra.
Since the coordinates commute or anticommute, their Jacobi
identities, involving three $Y$'s, are trivially satisfied.
Similarly, the super Jacobi identities involving two $Y$'s and 
one $X$ are also trivially satisfied.

We first consider the Jacobi identities involving three $X$'s. 
For the vector fields $X(a,\dot a),$ $X(b, \dot b)$ and  $X(c, \dot c)$, 
all three double-supercommutators yield linear combinations
of vector fields 
$X(f,\dot f)$, with
$(f,\dot f)$ belonging to the set obtained in the decomposition of
$(a\otimes b\otimes c, \dot a\otimes \dot b\otimes \dot c)$,
multiplied by products of $\epsilon$'s (realising the Clebsch-Gordan
coefficients) and quadratic in the $t$'s. They have terms of the form 
\be
     t(a,\dot a\,;b,\dot b\,;d,\dot d) \ t(d,\dot d\,;c,\dot c\,;f,\dot f)
\ S[A]S[\dot A]S[B]S[\dot B] S[C]S[\dot C] 
\e_{..}\ \ldots \ \e_{..}  X^{[\dot F]}_{[F]}\ .
\ee
The vanishing of the coefficients of the linearly independent tensors
$\e_{..}\ \ldots \ \e_{..}  X^{[\dot F]}_{[F]}$, 
are quadratic consistency conditions for the $t$ structure constants, 
which we consider to
be the defining relations for algebras $\A$. 
In constructing the linearly independent tensors, Fierz-type
identities based on the spinorial identity 
$ \e_{\a\b}T_\g  + \e_{\b\g}T_\a + \e_{\g\a}T_\b =0$
need to be taken into account.
For each $a,\dot a,b,\dot b,c,\dot c,f,\dot f$ and
for each
linearly independent tensor, we obtain a relation of the form
\bea
       \sum_{d,\dot d} &&\left(   
R_1(d,\dot d)\ t(a,\dot a\,;b,\dot b\,;d,\dot d)\ 
     t(d,\dot d\,;c,\dot c\,;f,\dot f)\right.
       \nonumber\\
                &&\quad +(-1)^{p_2}\ 
R_2(d,\dot d)\  t(b,\dot b\,;c,\dot c\,;d,\dot d)\ 
       t(d,\dot d\,;a,\dot a\,;f,\dot f)
        \nonumber\\[7pt]
               &&\quad +(-1)^{p_3} \ \left.  
R_3(d,\dot d)\  t(c,\dot c\,;a,\dot a\,;d,\dot d)\ 
          t(d,\dot d\,;b,\dot b\,;f,\dot f)
                     \right)=0\ ,
\label{jacobitt}
\eea
where the numerical constants $R_i$ $(i= 1,2,3)$ depend on the 
relation involved and where 
the range of the summation over $d,\dot d$ is given by 
representations occurring in 
the tensor products $(a,\dot a)\otimes (b,\dot b)$,
$(b,\dot b)\otimes (c,\dot c)$ and $(c,\dot c)\otimes (a,\dot a)$,
for the three terms respectively.
Here, $p_2=4(a+\dot a)(b+\dot b+c+\dot c)$ mod 2 is the parity 
of the permutation 
from $(a,\dot a\,;b,\dot b\,;c,\dot c)$ to 
$(b,\dot b\,;c,\dot c\,;a,\dot a)$ while
$p_3=4(c+\dot c)(a+\dot a+b+\dot b)$ mod 2
is the parity of the permutation 
from $(a,\dot a\,;b,\dot b\,;c,\dot c)$ to 
$(c,\dot c\,;a,\dot a\,;b,\dot b)$. For three bosons, three
fermions, or two bosons and one fermion in 
$(a,\dot a\,;b,\dot b\,;c,\dot c)$ this parity is always $0$. For two
fermions and one boson, one of them is $0$ and the other is $1$.

In practice and for simple examples, the linearly independent tensors
and the corresponding constant coefficients $R_i$
afford direct determination, using for instance REDUCE. In more
generality, a more precise form of \r{jacobitt} may be obtained using
$6j$ coefficients. Denoting the individual eigenstates of a Lorentz 
representation $T(a,\dot a)$ by
\be
T(a,a_3\,;\dot a,\dot a_3)  \ \ \ -a\leq a_3\leq a\ \ ,
                        \ \ \ -\dot a\leq \dot a_3\leq \dot a\ ,
\ee
where $a_3$, $\dot a_3$ label the eigenstates,
the super commutation relations between say  
$X(a,a_3\,;\dot a,\dot a_3)$ and $X(b,b_3\,;\dot b,\dot b_3)$ then read
\bea
&& \left[ X(a,a_3\,;\dot a,\dot a_3),
X(b,b_3\,;\dot b,\dot b_3) \right] 
                 \nonumber  \\
&&= \sum_{c=\mid a-b\mid}^{a+b}
       \sum_{\dot c=\mid \dot a-\dot b\mid}^{\dot a+\dot b}
    \wt t(a,\dot a\,;b,\dot b\,;c,\dot c)  
                 \nonumber\\
&&       C(a,a_3,b,b_3\,;c,a_3+b_3)
       C(\dot a,\dot a_3,\dot b,\dot b_3\,;\dot c,\dot a_3+\dot b_3)
       X(c,a_3+b_3\,;\dot c,\dot a_3+\dot b_3)
\label{comXXs} 
\eea 
where $C(a,a_3,b,b_3\,;c,a_3+b_3)$ is the $su(2)$ Clebsch-Gordan
coefficient coupling the state $(a,a_3)$ with the state
$(b,b_3)$ to form the state $(c,c_3)$ (where $c_3=a_3+b_3$ and 
analogously for the dotted indices) 
and the $\wt t$'s are renormalised $t$'s.
This is an alternative form of the commutation relations \r{comXX}.
Now, the $6j$ recoupling coefficients are defined by the relations
\bea
  &&C(a,a_3,b,b_3\,;d,a_3+b_3)C(d,a_3+b3,c,c_3\,;f,a_3+b_3+c3)
\nonumber\\
  &&=\sum_{k}R(a,b,c,d,k,f)
C(a,a_3,c,c_3\,;k,a_3+c_3)C(k,a_3+c_3,b,b_3\,;f,a_3+b_3+c3).
\label{6j}
\eea
If the Clebsch-Gordan coefficient has the symmetry
\be
C(a,a_3,b,b_3\,;d,d_3)=C(b,b_3,a,a_3\,;d,d_3)
\label{symcb}
\ee
then the $6j$ symbol satisfies
\be
R(a,b,c,d,k,f)=R(b,a,c,d,k,f)\ .
\label{sym6j}
\ee
The super-Jacobi identities are given by
\bea
        &&  
\wt t(a,\dot a\,;b,\dot b\,;d,\dot d)
    \ \wt t(d,\dot d\,;c,\dot c\,;f,\dot f)
  \label{jacobitts}\\
                &+&(-1)^{p_2}
\sum_{k,\dot k}
               R(b,c,a,k,d,f)
               R(\dot b,\dot c,\dot a,\dot k,\dot d,\dot f)
       \, \wt t(b,\dot b\,;c,\dot c\,;k,\dot k)\ 
       \, \wt t(k,\dot k\,;a,\dot a\,;f,\dot f)
      \nonumber\\
               &+&(-1)^{p_3}  
\sum_{k,\dot k}
               R(a,c,b,k,d,f)
              R(\dot a,\dot c,\dot b,\dot k,\dot d,\dot f)
       \, \wt t(c,\dot c\,;a,\dot a\,;k,\dot k)\ 
       \, \wt t(k,\dot k\,;b,\dot b\,;f,\dot f)
                     =0.
\nonumber
\eea
For given $a,\dot a,b,\dot b,c,\dot c$, i.e. the
starting three $X$ operators and the final $X$ operator $f,\dot
f$ there are as many relations as there are allowed $d,\dot d$ sets. 
This is the precise form of the more schematic relations \r{jacobitt}.

The Jacobi identities involving two $X$'s and one $Y$, say
$X(a,\dot a), X(b, \dot b)$ and $Y(c, \dot c)$,
imply two classes of conditions. One class involves the structure
constants $t$ and $u$ and is linear but inhomogeneous in $t$ and 
the other involves $t$, $u$ and $c$ and is strictly linear in $c$. 
Amongst $t,u$ we obtain, for each linearly independent combination of
tensors for the $Y$'s, a condition of the form
\bea
   \sum_{d,\dot d} &&\left( 
 S_1(d,\dot d)\ t(a,\dot a\,;b,\dot b\,;d,\dot d)\ 
     u(d,\dot d\,;c,\dot c\,;f,\dot f)\right.
       \nonumber\\
&&\quad -S_2(d,\dot d)\ u(b,\dot b\,;c,\dot c\,;d,\dot d)
    \ u(a,\dot a\,;d,\dot d\,;f,\dot f)
       \nonumber\\[7pt]
&&\quad + (-1)^{q_3}\ S_3(d,\dot d)\ \left. u(a,\dot a\,;c,\dot c\,;d,\dot d)
             \ u(b,\dot b\,;d,\dot d\,;f,\dot f)
              \right) = 0
\label{jacobitu}
\eea
where the numerical coefficients $S_i$ $(i= 1,2,3)$ depend on the 
relation involved. Here 
$q_3=4(a+\dot a)(b+\dot b)$ mod 2 is the parity of
the permutation from 
$(a,\dot a\,;b,\dot b)$ to  $(b,\dot b\,;a,\dot a)$ 
and the allowed values of the summation
indices $(d,\dot d)$ are the same as in \r{jacobitt}.

Finally, between $t,u$ and $c$ we obtain the conditions
\bea
    &&  T_1\  t(a,\dot a\,;b,\dot b\,;c,\dot c)\ c(c,\dot c)
    \nonumber\\
  &-& T_2\  u(b,\dot b\,;c,\dot c\,;a,\dot a)\ c(a,\dot a)
     \nonumber\\
  &+& (-1)^{q_3} T_3\  u(a,\dot a\,;c,\dot c\,;b,\dot b)\ c(b,\dot b)
        =0\ ,
\label{jacobitc}
\eea
with the numerical coefficients $T_i$ $(i= 1,2,3)$.
A more explicit form of relations (\ref{jacobitu}) and (\ref{jacobitc}),
in terms of $6j$ coefficients,
may be derived analogously to \r{jacobitts}.

Conditions (\ref{jacobitt}),(\ref{jacobitu}),(\ref{jacobitc}) are the
only restrictions amongst the structure constants arising from the 
super Jacobi identities. 
If some representations occur multiply, then these conditions of course
need to be modified, in order to accommodate the extra labelling indices 
of $t,u$ and $c$, as described in appendix \ref{mult}.  

\section{Explicit examples of solutions to the Jacobi identities}

In this section we present some simple examples of solutions to 
the conditions \r{jacobitt},\r{jacobitu},\r{jacobitc} and thereby
provide explicit examples of hyperspaces $\M$.  

Two natural, though by no means necessary, assumptions are:

\noindent
(a) that the Lorentz generators $M_{\alpha\beta}$
and $M^{\dot\alpha\dot\beta}$ are identical to the 
generators 
$X(1,0)$ and $X(0,1)$ respectively; and

\noindent
(b) that $X(0,0)$, which is basically a dilatation-type operator, 
and its corresponding coordinate $Y(0,0)$ are absent.

\noindent
The latter may be implemented in the relations for the structure
constants (\ref{jacobitt}),(\ref{jacobitu}) and (\ref{jacobitc}) 
either by ignoring the two operators from the beginning or equivalently
by imposing, for all $a,\dot a, b,\dot b$, the constraints,
\be\arr
&&t(0,0\,;a,\dot a\,;b,\dot b)\ =\ t(a,\dot a\,;b,\dot b\,;0,0)
\ =\  c(0,0)\ =\ 0,\\[5pt]
&&u(0,0\,;a,\dot a\,;b,\dot b)\ =\ u(a,\dot a\,;0,0\,;b,\dot b)
\ =\  u(a,\dot a\,;b,\dot b\,;0,0)\ =\ 0\  .
\ea\la{00}\ee
Using (\ref{lorentz2}), assumption (a) 
fixes unambiguously the following structure constants, for
all $a,\dot a, b,\dot b$,
\be\begin{array}{rcl} 
&&t(a,\dot a\,; 0,1\,; b,\dot b) \ =\  t(a,\dot a\,; 1,0\,; b,\dot b)
\ =\  t(0,1\,;a,\dot a\,;b,\dot b) \ =\  t(1,0\,;a,\dot a\,;b,\dot b) \ =\ 
\d_{ab}\d_{\dot a \dot b}\ ,\\[5pt]
&& u(0,1\,; a,\dot a\,;b,\dot b) \ =\   u(1,0\,; a,\dot a\,;b,\dot b)\ =\ 
\d_{ab}\d_{\dot a \dot b}\ ,\\[5pt]
&&c(0,1)\ =\ c(1,0) \ =\   0 \ .
\ea\la{02}\ee
A further natural constraint consistent with (a) is that
\be t(a,\dot a\,;b,\dot b\,;0,1) = \d_{a0}\d_{\dot a 1}\d_{b0}\d_{\dot b 1}\ 
 ,\quad 
t(a,\dot a\,;b,\dot b\,;1,0) = \d_{a1}\d_{\dot a 0}\d_{b1}\d_{\dot b 0}\ .
\la{lor}\ee

\subsection{}
Of course, any number of abelian examples of $\A$ may be constructed,
with arbitrary set of vector fields $\left\{  X\right\} $, representable in 
a corresponding coordinate basis $\left\{  Y\right\} $ by partial derivatives
\r{partial}. All $t$'s and $u$'s are then zero; and all $c$'s are 1,
apart from the vanishing $c(0,1)$ and $c(1,0)$. 

\subsection{}
If, apart from the Lorentz generators, the set of operators is
restricted to $\left\{  X(\half,0), X(0,\half),\right. $ 
$\left. X(\half,\half ), Y(\half,0), Y(0,\half), Y(\half,\half)\right\} $ 
and the corresponding set of
nonzero structure constants restricted to
$\left\{  t(0,\half\,;\half,0\,;\half,\half),\right. $ 
$u(0,\half\,;\half,\half\,;\half,0),$ 
$u(\half,0\,;\half,\half\,;0,\half),$ 
$\left. c(\half,\half), c(0,\half), c(\half,0)\right\} $
together with those in \r{lor} and \r{02}, the associativity
requirements of the previous section correspond to a single relationship
amongst the non-zero structure constants, viz.,
\be
t(0,\half\,;\half,0\,;\half,\half)\ c(\half,\half)=
c(0,\half)\ u(\half,0\,;\half,\half\,;0,\half)
+c(\half,0)\ u(0,\half\,;\half,\half\,;\half,0)
\ .
\la{sp}\ee
If we choose $c(\half,\half)=c(0,\half)=c(\half,0)=1$, we obtain a simple
relation for the determination of a coordinate basis consistent with 
the super-Poincar\'e relation (\ref{st1}).
 
\subsubsection{}
One explicit solution 
\be
t(0,\half\,;\half,0\,;\half,\half)= 2i,\quad 
u(\half,0\,;\half,\half\,;0,\half) = u(0,\half\,;\half,\half\,;\half,0) =i\ ,
\ee
corresponds to the standard super Poincar\'e basis (\ref{st3}) with
relations \r{st2} and is {\it{maximally non-chiral}} \r{adasymmetry}.

\subsubsection{}
Another solution
\be
t(0,\half\,;\half,0\,;\half,\half)= 2i,\quad 
u(\half,0\,;\half,\half\,;0,\half) = 2i,\quad 
u(0,\half\,;\half,\half\,;\half,0) =0\ ,
\ee
corresponds to the {\em chiral basis} for super Poincar\'e space
\be  
X_{\a\db} = \der{Y^{\a\db}} ,\quad 
X_{\a}= \der{Y^\a} + 2iY^\db \der{Y^{\a\db}} ,\quad 
X_{\db}= \der{Y^\db}\  .\la{chiral}
\ee
\subsubsection{}
The general solution to \r{sp} clearly interpolates between these two
bases for the super Poincar\'e algebra:
\be
 t(0,\half\,;\half,0\,;\half,\half)= 2i,\quad 
 u(\half,0\,;\half,\half\,;0,\half) = i(1+r),\quad 
 u(0,\half\,;\half,\half\,;\half,0) =i(1-r)\ ,
\ee
and can be realised as
\be  
X_{\a\db} = \der{Y^{\a\db}} ,\quad 
X_{\a}= \der{Y^\a} + i(1+r)Y^\db \der{Y^{\a\db}} ,\quad 
X_{\db}= \der{Y^\db} + i(1-r)Y^\a \der{Y^{\a\db}}\  .
\la{interpol}
\ee

\subsection{}
To find a simple example of a non-trivial extension of the
super Poincar\'e algebra, we have written, using
REDUCE, the set of all the conditions when we start with the set
of all $X$'s and $Y$'s 
transforming according to $(a,\dot a)$ representations, with 
$a+\dot a\leq 3$. Even for this simplest extension of the \sp, the
number of algebraic relations among the structure constants is rather
large. Using REDUCE, we find, for all representations with $a+\dot a\leq 3$,
a total of 397 relations of the form \r{jacobitt}, 1224 relations of the 
form \r{jacobitu} and 61 relations of the form \r{jacobitc},
which we call, respectively, TT, TU and TC relations.
The complete discussion of all allowed possibilities is a formidable task, 
however, the imposition of certain natural requirements yields a simple 
specific solution depending on a small number of arbitrary parameters.
Imposing 
\renewcommand{\theenumi}{\roman{enumi}}
\begin{enumerate}
\item   the absence of $X(0,0)$ and $Y(0,0)$, i.e. \r{00},
\item   the Lorentz identifications 
$X_{\a\b}=M_{\a\b},\quad X^{\da\db}=M^{\da\db}$ 
tantamount to \r{02}, together with \r{lor},   
\item   the condition that the essentially dummy variables 
$Y(1,0)$ and $Y(0,1)$ are zero, 
\item  the normalisations: 
$$ c(0,\half)=c(\half,0)=c(\half,\half)=c(1,\half)=c(\half,1)=
c(0,\3h)=c(\3h,0) =1\ ,
$$ 
\end{enumerate}
considerably reduces the number of relations to be satisfied by the
non-zero structure constants, though this is still quite large. However, 
if we in addition insist on the following super-Poincar\'e properties 
\begin{enumerate}\setcounter{enumi}{4}
\item  $t(0,\half\,;\half,0\,;\half,\half) \neq 0 $, 
\item  $u(\half,0\,;0,\half\,;\half,\half)
= u(0,\half\,;\half,0\,;\half,\half)=0$,
\item $u(\half,\half\,; a,\dot a\,;b,\dot b) = 0\  ,\quad
\mbox{for all}\ (b,\dot b)
\  \mbox{when}\ 2(a+\dot a)\ \mbox{is\ odd}$,
\end{enumerate}
then the TT-relations imply that $X(\half,\half)$ necessarily 
commutes with all fermionic $X$'s,
\begin{enumerate}\setcounter{enumi}{7}
\item
$t(\half,\half\,; a,\dot a\,;b,\dot b) = 0\  ,\quad
\mbox{for all}\ (b,\dot b)
\  \mbox{when}\ 2(a+\dot a)\ \mbox{is\ odd}$,
\end{enumerate}
as a consequence of which, all further TT-relations are automatically 
satisfied. Similarly, the TC-relations imply that
\begin{enumerate}\setcounter{enumi}{8} 
\item $ u(a,\dot a\,;b,\dot b\,;\half,\half) = 0\  ,\quad
\mbox{for all}\ (b,\dot b)
\  \mbox{when}\ 2(a+\dot a)\ \mbox{is\ odd}$,
\end{enumerate}
as a consequence of which all TU-relations are resolved. The only conditions
then remaining are the following TC-relations amongst the non-zero 
structure constants (all other structure constants being zero, except of
course for the non-zero ones in \r{02}, \r{lor})
\be\arr
t(0,\half\,;\half,0\,;\half,\half)&=& 
u(\half,0\,;\half,\half\,;0,\half)+u(0,\half\,;\half,\half\,;\half,0) \\[5pt]
t(\half,0\,;1,\half\,;\half,\half)&=& 
u(1,\half\,;\half,\half\,;\half,0)-u(\half,0\,;\half,\half\,;1,\half)\\[5pt]
t(0,\half\,;\half,1\,;\half,\half)&=& 
u(\half,1\,;\half,\half\,;0,\half)-u(0,\half\,;\half,\half\,;\half,1)\\[5pt]
t(\half,1\,;1,\half\,;\half,\half)&=& 
u(1,\half\,;\half,\half\,;\half,1)+u(\half,1\,;\half,\half\,;1,\half)\\[5pt]
t(1,\half\,;\3h,0\,;\half,\half)&=& 
u(1,\half\,;\half,\half\,;\3h ,0)+u(\3h,0\,;\half,\half\,;1,\half)\\[5pt]
t(0,\3h\,;\half,1\,;\half,\half)&=& 
u(\half,1\,;\half,\half\,;0,\3h)+u(0,\3h\,;\half,\half\,;\half,1) \ .
\ea\ee
Denoting the 12 arbitrary parameters 
\be\begin{array}{rllll}
u(\half,0\,;\half,\half\,;0,\half)= u_1,&& 
u(1,\half\,;\half,\half\,;\half,0)= u_2,&&
u(\half,0\,;\half,\half\,;1,\half)= u_3,\\[5pt]
u(0,\half\,;\half,\half\,;\half,0)=\wt u_1 ,&&
u(\half,1\,;\half,\half\,;0,\half)=\wt u_2\ ,&&
u(0,\half\,;\half,\half\,;\half,1)= \wt u_3,  \\[5pt]
u(1,\half\,;\half,\half\,;\3h ,0)=u_4,&&
u(\3h,0\,;\half,\half\,;1,\half)= u_5,&&
u(1,\half\,;\half,\half\,;\half,1)=u_6,\\[5pt]
u(\half,1\,;\half,\half\,;0,\3h)=\wt u_4,&&
u(0,\3h\,;\half,\half\,;\half,1)= \wt u_5,&&
u(\half,1\,;\half,\half\,;1,\half)=\wt u_6 \ ,
\ea\ee
we obtain a super algebra with non-zero commutation relations among the 
$X$'s (apart from the obvious relations involving the Lorentz generators
$X(1,0)$ and $X(0,1)$),
\begin{equation}
\begin{array}{rll}
\left\{  X_{\a} , X^{\db}\right\}  &=&  (u_1+\wt u_1)\  X_{\a}^{\db}
                         \\[8pt]
\left\{  X_{\a} , X_{\b_1\b_2}^{\db}\right\}  &=&  
(u_2- u_3)\ S(\b_1\b_2) \e_{\a\b_1}X_{\b_2}^{\db}
                         \\[8pt]
\left\{  X^{\da} , X_{\b}^{\db_1\db_2}\right\}  &=&  
(\wt u_2- \wt u_3)\ S(\db_1\db_2) \e^{\da\db_1}X_{\b}^{\db_2}
                        \\[8pt]
\left\{  X_{\a_1\a_2}^{\da} , X_{\b}^{\db_1\db_2}\right\}  &=&  
( u_6 + \wt u_6)\ S(\a_1\a_2) S(\db_1\db_2) 
\e_{\a_1\b}\e^{\da\db_1}X_{\a_2}^{\db_2}
                          \\[8pt]
\left\{  X_{\a_1\a_2\a_3} , X_{\b_1\b_2}^{\db}\right\}  &=&  
( u_4 + 2 u_5)\ S(\a_1\a_2\a_3) 
\e_{\a_1\b_1}\e_{\a_2\b_2}X_{\a_3}^{\db}
                          \\[8pt]
\left\{  X_{\a}^{\da_1\da_2} , X^{\db_1\db_2\db_3}\right\}  &=&  
( \wt u_4 + 2\wt u_5)\  S(\db_1\db_2\db_3) 
\e^{\da_1\db_1}\e^{\da_2\db_2}X_{\a}^{\db_3}\ .
\label{casbecxx}
\end{array}
\end{equation}
This algebra clearly has the symmetry corresponding to
the $\Z_2$ `chiral' transformations \r{tildesym}
which in this case reduce to
\be\arr
{\mbox{dotted upper index}} 
              &\leftrightarrow & {\mbox{undotted lower index}} 
                 \\[5pt] 
{\mbox{dotted lower index}} 
              &\leftrightarrow & {\mbox{undotted upper index}} 
                 \\[5pt] 
u_i&\leftrightarrow & \wt u_i
\ea\label{tildesym1}
\ee 
 
The non-zero commutators between the $X$'s
and the $Y$'s are:
\begin{equation}
\begin{array}{rll}
\left\{  X_{\a} , Y_{\b } \right\}  &=& 
\e_{\a\b}\ 
                            \\[8pt]
\left[  X_{\a}^{\da} , Y_{\b}^{\db}\right] & =&  
\e_{\a\b}\e^{\da\db} 
                          \\[8pt]  
\left\{   X_{\a_1\a_2}^{\da} , Y_{\b_1\b_2}^{\db} \right\}  &=& 
S(\a_1\a_2) \e_{\a_1\b_1} \e_{\a_2\b_2} \e^{\da\db} 
                          \\[8pt] 
\left\{   X_{\a_1\a_2\a_3} , Y_{\b_1\b_2\b_3} \right\}  &=& 
S(\a_1\a_2\a_3) 
\e_{\a_1\b_1} \e_{\a_2\b_2} \e_{\a_3\b_3} 
                          \\[12pt]
\left[  X_{\a} , Y_{\b}^{\db}\right]  &=& 
u_1\ \e_{\a\b}Y^{\db} + u_3\ Y_{\a\b}^{\db}
                          \\[8pt]
\left[  X_{\a_1\a_2}^{\da} , Y_{\b}^{\db}\right]  &=& 
u_2 S(\a_1\a_2) \e_{\a_1\b} \e^{\da\db} Y_{\a_2}  + 
u_4 \e^{\da\db} Y_{\a_1\a_2\b} + 
u_6 S(\a_1\a_2) \e_{\a_1\b} Y_{\a_2}^{\da\db}
                          \\[8pt]
\left[  X_{\a_1\a_2\a_3} , Y_{\b}^{\db}\right]  &=& 
 u_5\ S(\a_1\a_2\a_3) \e_{\a_1\b} Y_{\a_2\a_3}^{\db}
                    \ \ .
\end{array}
\label{casbecxy}
\end{equation}
together with the six further independent relations obtained 
by performing the transformations \r{tildesym1}.

A coordinate representation of this algebra may be found in
terms of a holonomic basis of vector fields
$\p_{[A]}^{\dot{[A]}}= 
S[A]S[\dot A] \partial / \partial Y_{[\dot A]}^{[A]}$, 
having commutation relations corresponding to
(\ref{casbecxx}) and (\ref{casbecxy}) with all the $u$'s put to
zero. This representation is given by
\begin{equation}
\begin{array}{rll}
X_{\a}^{\da} &=& \p_{\a}^{\da}  \\
X_\a &=& \p_\a   
            + u_1\ \e_{\da\db}Y^\da\ \p_{\a}^{\db}              
            + u_3\ \e_{\da\db}\e^{\b\g}Y_{\a\b}^{\da}\ \p_{\g}^{\db} \\
X_{\a\b}^{\da} &=& \p_{\a\b}^{\da} 
            + u_2\ S(\a\b)\ Y_{\a}\ \p_{\b}^{\da}
            + u_4\ \e^{\g\d}Y_{\a\b\g}\ \p_{\d}^{\da} 
            + u_6\  S(\a\b)\ \e_{\db\dg}Y_{\a}^{\da\db}\  \p_{\b}^{\dg} \\
X_{\a\b\g} &=&  \p_{\a\b\g} 
        + u_5\ S(\a\b\g)\ \e_{\da\db}Y_{\a\b}^{\da}\ \p_{\g}^{\db} 
\end{array}
\end{equation}
with realisations for the remaining generators, 
$X^\da , X^{\da\db}_\a , X^{\da\db\dg}$, 
being given by the latter three expressions on performing transformations
\r{tildesym1}.

This example clearly has the super Poincar\'e algebra as a subalgebra. 

\section{Hyperspace Gauge Fields}

We would like to consider gauge fields on the spaces $\M$. 
To this end we postulate gauge potentials $A$.
These depend on the set of coordinates $\left\{  Y\right\} $
and are in one-to-one correspondence with the hypercovariant derivatives 
$X$, whose Lorentz indices and corresponding transformation properties they
carry,
allowing the definition of gauge-covariant derivatives,
\be
{\cal{D}}_{[A]}^{[\dot A]}=X_{[A]}^{[\dot A]}
                           +A_{[A]}^{[\dot A]}
\ .
\label{covariantderiv}
\ee
The gauge potentials take values in some, for instance semi-simple,
Lie algebra, 
with generators, 
$\left\{   \lambda_k ;\ k=1,\ldots,N\right\} $, satisfying 
\be
\left [\lambda_k,\lambda_l\right ]=f_{kl}^{{\phantom{kl}}m}\lambda_m\ ,\quad  
{\rm{tr}}(\lambda_k\lambda_l)=\delta_{kl}\ ,
\label{Liegenerators}
\ee
where $f_{kl}^{{\phantom{kl}}m}$ are the structure constants.
We can expand the gauge potentials in this Lie algebra basis 
\be
A_{[A]}^{[\dot A]}=\sum_{k=1}^{N} A_{[A]}^{[\dot A],k}\lambda_k  
\ee
with coefficients being extractable thus:
\be 
A_{[A]}^{[\dot A],k}
    ={\rm{tr}}\left (A_{[A]}^{[\dot A]}\lambda_k\right ) \ .   
\label{potentialcomponents}
\ee
The covariant transformation law
\be
{\cal{D}}_{[A]}^{[\dot A]} \rightarrow  
{{\cal{D}}'}_{[A]}^{[\dot A]}=U {\cal{D}}_{[A]}^{[\dot A]}U^{-1} 
   \ \ \  ,
\label{gtD}
\ee
yields the standard inhomogeneous gauge transformations
\be
A_{[A]}^{[\dot A]} \rightarrow  
{A'}_{[A]}^{[\dot A]}=U A_{[A]}^{[\dot A]}U^{-1} 
            -  \left[ X_{[A]}^{[\dot A]}, U \right] U^{-1}
   \ \ \  ,
\label{gt}
\ee
where the (grassmann-even) gauge-group-valued function $U$ depends on 
the coordinates $Y$. 
The infinitesimal version ($U=1+\tau +O(\tau^2)$) of this transformation
\be
A_{[A]}^{[\dot A]} \rightarrow 
{A'}_{[A]}^{[\dot A]}= A_{[A]}^{[\dot A]} - 
\left [\D_{[A]}^{[\dot A]}~,~ \tau \right]
\ ,  \label{gtinf}
\ee
where the components of the gauge potential are linear in a dimensionless
coupling constant (absorbed into their definition), reveals them to have 
the same scaling dimension and the same bosonic or fermionic nature as 
the corresponding $X$'s.
Although these potentials have general spin, the gauge transformations remain completely analogous to the usual gauge 
transformations corresponding to a spin-one gauge degree of freedom, 
with lorentz-scalar transformation parameter $\tau$ taking values in the
gauge algebra. So although the potentials $A_{[A]}^{[\dot A]}$
have, in general, higher than spin-one content, no higher-spin 
gauge-invariances and no coupling constants apart from the 
Yang-Mills ones are introduced.

It is now natural to define generalised gauge fields (curvatures)
$\hat F$ (corresponding to the sets $[A],[\dot A]$ and $[B],[\dot B]$
by the equation
\bea 
 {\hat F}_{[A] [B]}^{ [\dot A][\dot B]} 
 =&&\left [{\cal{D}}_{[A]}^{[\dot A]},{\cal{D}}_{[B]}^{[\dot B]}\right ] 
\nonumber\\  &&  
   -\ \sum_{s=0}^{\min (2a,2b)}\sum_{\dot s=0}^{\min (2\dot a,2\dot b)}
  t(a,\dot a\,;b,\dot b\,;a+b-s,\dot a+\dot b-\dot s)
\nonumber\\  &&  \hskip 3.4 true cm
     S[A]S[\dot A]S[B]S[\dot B]\epsilon_{[\a_s\b_s]}
           \epsilon^{[\da_{\dot s}\db_{\dot s}]}
     {\cal{D}}^{[C(s)]}_{[\dot C(\dot s)]}\ .
\label{gaugefield}
 \eea 
On the right hand side the second term 
ensures that the fields $\hat F$ are free of differential
operators in a gauge-covariant manner. 
The thus defined curvatures are manifestly gauge-covariant 
\be 
 {\hat F}_{[A] [B]}^{[\dot A][\dot B]} \rightarrow
 {\hat F'}{}_{[A] [B]}^{[\dot A][\dot B]}
  =U {\hat F}_{[A] [B]}^{[\dot A][\dot B]} U^{-1}
\label{gtgaugefields} 
\ee
and decompose under the action of the Lorentz group into irreducible
Lorentz representations thus:
\be 
\hat F_{[A] [B]}^{ [\dot A][\dot B]}= 
\sum_{s=0}^{\min (2a,2b)}\sum_{\dot s=0}^{\min (2\dot a,2\dot b)}
	S[A]S[\dot A]S[B]S[\dot B] \epsilon_{[\a_s\b_s]}
              \epsilon^{[\da_{\dot s}\db_{\dot s}]}
		     F_{[C(s)]}^{ [\dot C(\dot s)]}\ .
\label{decompF}
\ee
The irreducible components $F_{[C(s)]}^{[\dot C(\dot s)]}$
(which obviously depend on $[A], [B],[\dot A ],[\dot B]$)
transforming according to the $(a+b-s , \dot a+\dot b-\dot s)$
Lorentz representations may be projected out by contracting
the curvatures $\hat F$ with the inverse 
epsilon tensors $\epsilon^{[\b_s\a_s]}$ 
and $\epsilon_{[\db_{\dot s}\da_{\dot s}]}$ in
(\ref{epinv}) and symmetrizing over the remaining multi-indices 
$[C(s)]$ and $[\dot C(\dot s)]$ in (\ref{Csets}): 
\begin{equation} 
  F_{[C(s)]}^{[\dot C(\dot s)]} =  \kappa(s,\dot s)
  S[C(s)]S[\dot C(\dot s)] \epsilon^{[\b_s\a_s]}
            \epsilon_{[\db_{\dot s}\da_{\dot s}]} 
      \hat F_{[A] [B]}^{[\dot A ][\dot B]}
\ \  ,\label{irgaugefields}\end{equation}
where $\kappa(s,\dot s)$ are combinatorial factors.
The gauge algebra components of the irreducible fields $F$ may be
extracted by taking traces as in (\ref{potentialcomponents}).  

For instance, corresponding to the first two commutation relations in
\r{casbecxx}, we have
\be
\hat F_{\a\db} = \left\{  \D_{\a} , \D_{\db}\right\}  - 
(u_1+\wt u_1)\  \D_{\a\db} =
F_{\a\db}\ ,
\ee
which is irreducible and
\be
\arr
\hat F_{\a\b_1\b_2\db} &=& 
     \left\{  \D_{\a} , \D_{\b_1\b_2\db}\right\}  -  
(u_2- u_3)\ S(\b_1\b_2) \e_{\a\b_1}\D_{\b_2\db} \\[5pt]
&=&  S(\a\b_1\b_2) F_{\a\b_1\b_2\db} 
    + S(\b_1\b_2) \e_{\a\b_1} F_{\b_2\db} \ .
\ea
\ee

\section{Curvature constraints and integrability conditions}
\la{constr}
Having the basic ingredients of the previous section, we 
could now proceed to consider the possibility of constructing 
consistent lagrangian gauge field theories on the hyperspaces $\M$. 
We leave this for future work, concentrating here
on systems of equations for gauge fields on $\M$, which generalise
the idea of the standard four-dimensional self-duality equations.
We shall consider imposing `coherent' curvature constraints,
setting some sets of irreducible gauge fields to vanish. 
One important class of such constraints arises from by demanding that
some subset of the commutation relations of the superalgebra $\A\ $
are preserved under the covariantisation $X \rightarrow \D$.
This yields relations for higher-spin potentials in terms of lower-spin
ones. The simplest example is the `conventional constraint' of standard 
super Yang-Mills theory, 
\be
F_{\a\db}=\D_{\a} A_{\db} + \D_{\db} A_{\a} - (u_1+\wt u_1)\ A_{\a\db} =0\ ,
\ee
which determines the vector potential $A_{\a\da}$ in terms of the spinor 
ones $(A_\a, A_\da)$ and yields an irreducible representation of the 
supersymmetry algebra. 
Another class of constraints are
generalisations of the
standard four-dimensional self-duality equations. In particular,
the \sdy\ conditions are integrability conditions for a linear system,
the starting point for the twistor transform, which establishes
formal solubility. The crucial feature is the
possibility of writing the constraints, with the non-zero curvature 
components representing  obstructions to Frobenius integrability, 
in the form of some set of commuting operators.

In order to define linear systems we introduce
two commuting spinors, $v^{\op\a}$ and $u^{+\da}$.
Our notation for these spinors is 
explained in appendix \ref{harmapp}.
Let us take a set of representations $\{(a,\dot a)\}$ of the
Lorentz group and 
associate to each representation two freely specifiable 
integers:  $r(a),\ 0\leq r(a)\leq 2a$ and 
$\dot r(\dot a),\ 0\leq \dot r(\dot a)\leq 2\dot a$.    
We consider the linear systems  
\be
 v^{\op\a_1}\dots v^{\op\a_{r(a)}} u^{+\da_1}\dots u^{+\da_{\dot r(\dot a)}} 
       \D_{\a_1\dots \a_{2a}\da_1\dots \da_{2\dot a}} \f =\ 0.   
\la{ls}
\ee
When $r(a)=0$
(resp. $\dot r(\dot a)= 0$), we mean that $v$'s (resp. $u$'s) do not appear.

The integrability of these linear systems for {\it{arbitrary choices}}
of $u$'s and $v$'s is equivalent to Lorentz covariant curvature
constraints. 
Now, for field theories in
dimensions greater than two, there is no clear--cut definition of
`complete integrability', but for the cases where this notion has any 
meaning, four dimensional \sdy\ being the paradigm, the existence of a 
linear system is crucial.
In particular, it is central to solution generating transforms.

Of particular significance
amongst the systems \r{ls}, are those with  
$r(a)$ taking values $0$ or $2a$ only and  $\dot r(\dot a)$ taking values
$0$ or $2\dot a$ only. In other words, if, for
any $(a,\dot a)$, $u$ or $v$-type spinors exist, they saturate {\it all}
the corresponding indices on  $\D_{[A][\dot A]}$. These linear systems
therefore correspond to representations with indices $(a,\dot a)$ falling into
four disjoint sets $\Lambda_{ij}, i,j=0,1$ 
(where $i=r(a)$ mod $2a$ and $j=\dot r(\dot a)$ mod $2\dot a$) 
\bea
\Lambda_{00}=& \left\{ (a, \dot a)\mid
\D_{\a_1\dots \a_{2a}\da_1\dots \da_{2\dot a}} \f = 0 
               \right\}& \label{set1}\\[5pt]
\Lambda_{01}=& \left\{ (a, \dot a)\mid \ 2\dot a\geq 1,\ 
u^{+\da_1}\dots u^{+\da_{2\dot a}} 
       \D_{\a_1\dots \a_{2a}\da_1\dots \da_{2\dot a}} \f = 0
               \right\}& \label{set2}\\[5pt]
\Lambda_{10}=& \left\{ (a, \dot a)\mid \ 2a\geq 1,\
v^{\op\a_1}\dots v^{\op\a_{2a}}  
       \D_{\a_1\dots \a_{2a}\da_1\dots \da_{2\dot a}} \f = 0
               \right\}&  \label{set3}\\[5pt]
\Lambda_{11}=& \left\{ (a, \dot a)\mid \ 2a\geq 1,\ \ 2\dot a\geq 1,\
v^{\op\a_1}\dots v^{\op\a_{2a}}  
u^{+\da_1}\dots u^{+\da_{2\dot a}} 
       \D_{\a_1\dots \a_{2a}\da_1\dots \da_{2\dot a}} \f = 0
               \right\}& \label{set4}
\eea
The constraints which arise as integrability conditions for 
these systems
may be formally solved following 
the generalised twistor construction given in the appendix of \c{w}.
The harmonic space description of solutions may also be given on
the lines of the construction 
reviewed for instance in \c{n3ym,guer,sdharm}. 
The integrability conditions for \r{set1}-\r{set4} therefore
provide classes of systems which 
are completely solvable in the sense of the standard \sdy\ equations
and generalise the systems of higher dimensional solvable gauge
field constraints classified in \c{w}. 
They fall into two broad classes: 
\vskip 5pt
\begin{description}
\item{I)} with either only $u$-type or only $v$-type spinors present,
which are exact analogues of \sd\ or respectively \asd\ systems 
on $\M$.
\item{II)}  with both $u$-type and $v$-type spinors present,
which are analogues of light-like integrable systems.
\end{description}
\vskip 5pt

\noindent {\bf{Class I. Generalised Self-Duality}}

For gauge fields on an arbitrary hyperspace $\M$, we define self-duality
as the condition that the only non-zero
irreducible curvatures are those appearing in (\ref{decompF}) with
a coefficient including at least one $\epsilon$-tensor with
dotted indices, i.e. those in the decomposition
\be 
\hat F_{[A] [B]}^{ [\dot A][\dot B]}= 
\sum_{s=0}^{\min (2a,2b)}\sum_{\dot s=1}^{\min (2\dot a,2\dot b)}
	S[A]S[\dot A]S[B]S[\dot B] \epsilon_{[\a_s\b_s]}
            \epsilon^{[\da_{\dot s}\db_{\dot s}]}
		     F_{[C(s)]}^{ [\dot C(\dot s)]}\ .
\label{decompF1}
\ee

The vanishing of the curvatures appearing in (\ref{decompF}) but not in
(\ref{decompF1}) is a natural generalisation of the self-duality equations. 

For a given superspace $\M$, with coordinates enumerated by some set
of Lorentz indices 
$\Lambda = \left\{   (a,\dot a) \right\} $, 
the vanishing of these curvatures are integrability conditions 
for linear systems containing
equations of the form (\ref{set1}) with $2\dot a=0$ and (\ref{set2}) only.
The prototypical examples are standard \sdy\ and super \sdy.
Obviously analogous {\it{anti}}-self-dual systems containing
equations of the form
(\ref{set1})  with $2 a=0$ and (\ref{set3}) only 
can also be considered. In
the following examples we restrict ourselves to the former
self-dual systems. 
\vskip 10pt 
{\bf Example I-1 }  $\Lambda_{01} = \left\{  (\half,\half)\right\} $, 
with commutative
algebra of the four components of $X_{\a\da}$, (i.e. $\A$ is the
Poincar\'e algebra). The linear system  ($r=0, \dot r=1$)
\be
u^{+\da}\D_{\a\da} \f =\ 0
\label{sdymls}
\ee
is precisely the Belavin-Zakharov-Ward linear system for the 
\sdy\ equations. The consistency conditions are
\be
\left[ u^{+\da}\D_{\a\da}, u^{+\db}\D_{\b\db} \right]
= u^{+\da} u^{+\db}\left[\D_{\a\da},\D_{\b\db} \right]   
=\e_{\a\b} u^{+\da} u^{+\db} F_{\da\db}
=  0\ \ \ .  
\label{zcsd}
\ee
If we insist
that these hold for any choice of $u$, which is tantamount to
requiring their Lorentz covariance, we obtain
\be
\left[ \D_{\a\da}, \D_{\b\db} \right]
=\e_{\da\db}  F_{\a\b}\quad\Leftrightarrow\quad
F_{\da\db} =  0\  
\ee
i.e. the \sdy\ constraints. 
\vskip 10pt 
{\bf Example I-2 }
 $\Lambda_{00} = \left\{  (\half,0)\right\} , 
\Lambda_{01} = \left\{  (0,\half),(\half,\half)\right\} $, with
 algebra  of the vector fields $(X_\a, X_{\da},X_{\a\da})$
having non-zero supercommutator, 
$\left\{   X_\a , X_{\da}\right\} = 2X_{\a\da}$,
(i.e. $\A$ is the super-Poincar\'e algebra). The system
\be
  \D_\a \f =\ 0 ,\quad
 u^{+\db}\D_{\db}\f =\ 0 ,\quad
 u^{+\db}\D_{\a \db}\f =\ 0 
\ee
implies the \ssdy\ constraints. 

N-extended supersymmetrisations, on superspaces with
odd coordinates having multiplicity N,  
$ \left\{   Y^{\a\da},  Y^\a_m, Y^{m\da}; m = 1,\dots N \right\} $, 
and vector fields 
$\left\{  X_{\a\da}  X^m_{\a}, X_{m\da}; m=1,\dots N \right\} $ 
satisfying the algebra
\be
\left\{  X^m_{\a} , X_{n\db}\right\}  =\  2\d^m_n  X_{\a\db}\  ,
\label{stn} 
\ee
with all other supercommutators amongst the $X$'s vanishing and
Jacobi-allowed $X,Y$ supercommutators.
 
They exist for arbitrary N \c{do} and the linear system
\be
  \D^m_\a \f =\ 0 ,
\quad  u^{+\db}\D_{m\db}\f =\ 0 ,
\quad  \  u^{+\db}\D_{\a \db}\f  =\ 0 
\la{nsdls}
\ee
implies the following curvature constraints equivalent to the
N-extended \ssdy\ equations.
\be
\begin{array}{rrlllll}
  & \left\{  \D^m_{\a},  \D^n_{\b} \right\} & = & 0  
    & \quad\Leftrightarrow\quad F^{mn}_{\a\b} &=& 0 \ ,\ F^{mn}=\ 0  \\[7pt]
  & \left\{  \D^m_{\a},  \D_{n\da} \right\} & = & 2\d^m_n \D_{\a\da} 
    & \quad\Leftrightarrow\quad  F^m_{n\a\da}&=&0 \\[7pt]
  & \left[  \D^m_{\a}, \D_{\b \db}  \right] & = &0 
    & \quad\Leftrightarrow\quad F^m_{\a\b\db} &=& 0\ ,\ F^m_{\db}=\ 0 \\[7pt] 
  & \left\{  \D_{m\da},  \D_{n\db} \right\} & =& \e_{\da\db}F_{mn} 
    & \quad\Leftrightarrow\quad F_{mn\da\db} &=& 0  \\[7pt]
  & \left[ \D_{m\da}, \D_{\b \db}  \right] & = & \e_{\da\db} F_{m\b}  
    & \quad\Leftrightarrow\quad F_{m\b\da\db} &=& 0 \\[7pt]
  & \left[ \D_{\a\da}, \D_{\b \db}  \right] & = &  \e_{\da\db} F_{\a\b} 
    & \quad\Leftrightarrow\quad F_{\da\db} &=& 0    
\la{ssdy}
\ea
\ee
where, by construction, $F_{mn}$ and $F^{mn}$ are distinct curvatures.
\vskip 10pt 
{\bf Example I-3 }
 $\Lambda_{01} = \left\{  (\half,{n\over 2})\right\} $,
for fixed odd integer $n > 1$, with commutative
algebra of the elements $X_{\a\da_1 \dots \da_n }$. 
In the bosonic space of dimension 2(n+1)  with coordinates
$Y^{\a \da_1 \dots \da_n }$, the irreducible parts of the gauge
curvature  are given by 
\bea
\left[  \D_{\a \da_1 \dots \da_n }, \D_{\b \db_1 \dots \db_n }  \right]  
    &&   \nonumber \\[5pt]  
 \quad\quad\quad = S(\da_1 \dots \da_n)&S(\db_1 \dots \db_n)
    &\left( 
       \e_{\a\b} \sum_{m=0}^{\frac{n-1}{2}} 
      \e_{[\da_{2m}\db_{2m}]}
            F_{\da_{2m+1} \dots \da_n \db_{2m+1} \dots \db_n}\right. 
     \nonumber\\[5pt]
 & & \left.+ \sum_{m=0}^{\frac{n-1}{2}}
      \e_{[\da_{2m+1}\db_{2m+1}]}
 F_{\a\b \da_{2m+2} \dots \da_n \db_{2m+2} \dots \db_n } \right)
\la{8c}
\eea
where at the upper limit in the second sum 
$F_{\a\b \da_{n+1} \dots \da_n \db_{n+1} \dots \db_n }$ simply
stands for $F_{\a\b}$.

The linear system \c{w} 
\be
   u^{+\da_1}\dots u^{+\da_n} 
       \D_{\a \da_1 \dots \da_n }\f =\ 0
\label{wardls}
\ee
yields the Lorentz covariant constraint 
\be
F_{\da_1 \dots \da_n \db_1 \dots \db_n} = 0
\la{sd8}
\ee
which Ward considered 
as an example of a soluble system generalising \sdy\  \c{w}. 

We note that the linear systems 
\be
   u^{+\da_1}\dots u^{+\da_{\dot r}} 
       \D_{\a \da_1 \dots \da_n }\f =\ 0\  ,\quad 
      {\rm{for\ some \ fixed }}\ \ {\dot r},\ \ 0<{\dot r}<n
\label{weakls}
\ee
yield integrability conditions with more of the 
irreducible pieces of the curvature in the decomposition \r{8c}
vanishing. These gauge field constraints are however not amenable to 
the twistor transform, the mapping to twistor space variables not 
being invertible for $0<s<n$. It is therefore unclear to what extent 
such equations, arising as integrability conditions for linear systems,
are actually `integrable'. They do not appear to be exactly solvable 
in the sense of the standard \sdy\ equations.

\vskip 10pt 
{\bf Example I-4 }
 $\Lambda_{01} = \left\{  ({n\over 2},\half)\right\} $,
for fixed odd integer $n > 1$. This example is clearly a mirror
image of the previous one.
The irreducible parts of the gauge
curvature are 
\bea
 \left[  \D_{ \a_1 \dots \a_n\da }, \D_{ \b_1 \dots \b_n \db}  \right]  
   &&  \nonumber \\[5pt]
 \quad\quad\quad = S(\a_1 \dots \a_n)&S(\b_1 \dots \b_n)
    &\left( 
       \e_{\da\db} \sum_{m=0}^{\frac{n-1}{2}} 
      \e_{[\a_{2m}\b_{2m}]}
            F_{\a_{2m+1} \dots \a_n \b_{2m+1} \dots \b_n}\right. 
       \nonumber\\[5pt]
 &&\left. + \sum_{m=0}^{\frac{n-1}{2}}
      \e_{[\a_{2m+1}\b_{2m+1}]}
 F_{ \a_{2m+2} \dots \a_n \b_{2m+2} \dots \b_n \da\db} \right).
\la{8d}
\eea

The linear system \c{w} 
\be
   u^{+\da} 
       \D_{ \a_1 \dots \a_n\da }\f =\ 0\ .
\label{wardls2}
\ee
yields the constraints 
\be
 \left[  \D_{ \a_1 \dots \a_n\da }, \D_{ \b_1 \dots \b_n \db}  \right]  
   =  S(\a_1 \dots \a_n)S(\b_1 \dots \b_n)
       \e_{\da\db} \sum_{m=0}^{\frac{n-1}{2}} 
      \e_{[\a_{2m}\b_{2m}]}
            F_{\a_{2m+1} \dots \a_n \b_{2m+1} \dots \b_n} 
\ee
which are equivalent to 
\be
F_{\da\db}=0
\ee
together with
\be
 F_{ \a_{2m} \dots \a_n \b_{2m} \dots \b_n \da\db}=0
,\quad{\rm{for\ all}}\quad m=1,\ldots,{\textstyle{n-1\over 2}}
\ \ .
\ee
\vskip 10pt 
{\bf Example I-5 }
This example provides self-duality conditions for 
the hyperspace $\M\ $ of section 4..3. We take   
$\Lambda_{00} = \left\{({1\over 2},0),({3\over 2},0)\right\}$ and
$\Lambda_{01} = \left\{(0,{1\over 2}),(\half,{1\over 2}),
(1,{1\over 2}),(\half,1),(0,{3\over 2}),\right\}$. 

The seven equations of the corresponding linear system 
have 28 consistency conditions which  
imply that all the curvatures vanish except for
19 curvatures appearing with a coefficient consisting of at
least one $\epsilon$ with dotted indices. 

\noindent {\bf{ Class II. Light-like Integrable Systems}}

Including $u$ and $v$ spinors simultaneously, we may consider 
linear systems of the form \r{set1}-\r{set4}, with all spinorial indices
saturated, 
\be
\arr
       \D \f &=& 0 
                     \\[5pt]
u^{+\da_1}\dots u^{+\da_{2\dot a}} 
       \D_{\da_1\dots \da_{2\dot a}} \f &=& 0 
\quad{\mbox{for}}\ \ 2\dot a \geq 1
                     \\[5pt]
v^{\op\a_1}\dots v^{\op\a_{2a}}  
       \D_{\a_1\dots \a_{2a}} \f &=& 0
\quad{\mbox{for}}\ \ 2 a \geq 1
                    \\[5pt]
v^{\op\a_1}\dots v^{\op\a_{2a}}  
u^{+\da_1}\dots u^{+\da_{2\dot a}} 
       \D_{\a_1\dots \a_{2a}\da_1\dots \da_{2\dot a}} \f &=& 0
\quad{\mbox{for}}\ \ 2 a \geq 1,\ 2 \dot a \geq 1 
                     \label{set5}\ .
\ea
\ee

Their consistency conditions express integrability
along certain lines in $\M$, analogues of `super null lines' in \Mi. 
These correspond to the vanishing of all curvatures which appear
in the decomposition \r{decompF} with a
coefficient including at least one $\e$-tensor (with either
dotted or undotted indices).
In other words, the only non-zero
irreducible curvatures are those appearing in
\be 
\hat F_{[A] [B]}^{ [\dot A][\dot B]}= 
\sum_{s=0}^{\min (2a,2b)}\sum_{\dot s=1}^{\min (2\dot a,2\dot b)}
	S[A]S[\dot A]S[B]S[\dot B] \epsilon_{[\a_s\b_s]}
             \epsilon^{[\da_{\dot s}\db_{\dot s}]}
		     F_{[C(s)]}^{ [\dot C(\dot s)]}
\label{decompF2}
\ee
or in
\be
\hat F_{[A] [B]}^{ [\dot A][\dot B]}= 
\sum_{s=1}^{\min (2a,2b)}\sum_{\dot s=0}^{\min (2\dot a,2\dot b)}
	S[A]S[\dot A]S[B]S[\dot B] \epsilon_{[\a_s\b_s]}
              \epsilon^{[\da_{\dot s}\db_{\dot s}]}
		     F_{[C(s)]}^{ [\dot C(\dot s)]}
\label{decompF3}
\ee
and the zero curvatures are those appearing in
(\ref{decompF}) but not in (\ref{decompF2}) or in (\ref{decompF3}).
\vskip 10pt 
{\bf Example II-1} The linear equations in standard Minkowski space 
\be
v^{\op\a}u^{+\db}\D_{\a \db}\f =\ 0 
\ee
clearly imply
\be
\left[ \D_{\a\da}, \D_{\b \db}  \right]  =   \e_{\a\b} F_{\da\db}
                                                +  \e_{\da\db} F_{\a\b}
\ee
and hence impose no constraints on the curvature.
More generally, this conclusion follows from \r{set5} whenever $\Lambda$
contains only one bosonic representation.   
\vskip 10pt 
{\bf Example II-2} The linear equations on $N$-extended
super Minkowski space corresponding to algebra \r{stn},
\be
  v^{\op\a}\D^m_\a  \f =\ 0,\quad
 u^{+\db}\D_{n \db}\f =\ 0,\quad v^{\op\a}u^{+\db}\D_{\a \db}\f =\ 0 
\ee
have as integrability conditions the conventional 
superspace constraints for N=1,2,3 \sym\ theories \c{s},
\be
\begin{array}{rrlllll}
  & \left\{  \D^m_{\a},  \D^n_{\b} \right\} & = & \e_{\a\b}F^{mn}  
    & \quad\Leftrightarrow\quad F^{mn}_{\a\b} &=& 0    \\[7pt]
  & \left\{  \D^m_{\a},  \D_{n\da} \right\} & = & 2\d^m_n \D_{\a\da} 
    & \quad\Leftrightarrow\quad  F^m_{n\a\da}&=&0 \\[7pt]
  & \left[  \D^m_{\a}, \D_{\b \db}  \right] & = &\e_{\a\b}F^m_{\db} 
    & \quad\Leftrightarrow\quad F^m_{\a\b\db} &=& 0 \\[7pt] 
  & \left\{  \D_{m\da},  \D_{n\db} \right\} & =& \e_{\da\db}F_{mn} 
    & \quad\Leftrightarrow\quad F_{mn\da\db} &=& 0  \\[7pt]
  & \left[ \D_{m\da}, \D_{\b \db}  \right] & = & \e_{\da\db} F_{m\b}  
    & \quad\Leftrightarrow\quad F_{m\b\da\db} &=& 0 \\[7pt]
  & \left[ \D_{\a\da}, \D_{\b \db}  \right] & = &  \e_{\a\b} F_{\da\db}
                                                +  \e_{\da\db} F_{\a\b}
    &                                         & & 
\ .   
\la{n3}
\ea
\ee

These are needed in order that the superfield carries an irreducible
representation of the supersymmetry algebra.
For N=1,2 these constraints do not have any dynamical consequences,
but for N=3 they turn out to be {\it equivalent} to the full
(second-order) N=3 \ym\  equations for the component fields 
in ordinary \Mi\  \c{s,hhls}.

For N=4, a further constraint is necessary to have an irreducible
supermultiplet, viz., 
\be
  F^{ij} =\ \half  \e^{ijkl} F_{kl} 
\ee
and these together with \r{n3} are similarly equivalent to the full
\sym\ equations for the N=4 \ym\ multiplet. This further constraint
does not arise as an integrability condition of the type considered 
here, so this is one example of an interesting set of constraints 
which is {\it not} a consequence of \r{ls}.
\vskip 10pt 
{\bf Example II-3 }
$\Lambda_{11} = \left\{  (\half,{n\over 2})\right\} $,
for fixed even integer $n > 1$. The  
curvatures in this purely fermionic space are given by 
\bea
\left\{  \D_{\a \da_1 \dots \da_n }, \D_{\b \db_1 \dots \db_n }  \right\}  
    &&   \nonumber \\[5pt]  
 \quad\quad\quad = S(\da_1 \dots \da_n)&S(\db_1 \dots \db_n)
    &\left( 
       \e_{\a\b} \sum_{m=0}^{\frac{n-2}{2}} 
      \e_{[\da_{2m+1}\db_{2m+1}]}
            F_{\da_{2m+2} \dots \da_n \db_{2m+2} \dots \db_n}\right. 
     \nonumber\\[5pt]
 & & \left.+ \sum_{m=0}^{\frac{n}{2}}
      \e_{[\da_{2m}\db_{2m}]}
 F_{\a\b \da_{2m+1} \dots \da_n \db_{2m+1} \dots \db_n } \right)
\la{8cII}
\eea
where at the upper limit in the second sum 
$F_{\a\b\da_{n+1} \dots \da_n \db_{n+1} \dots \db_n }$
stands for $F_{\a\b}$.

The linear system 
\be
   v^{\op\a}u^{+\da_1}\dots u^{+\da_n} 
       \D_{\a \da_1 \dots \da_n }\f =\ 0
\label{wlsII}
\ee
yields the Lorentz covariant constraint 
\be
F_{\a\b\da_1 \dots \da_n \db_1 \dots \db_n} = 0
\la{sd8II}
\ee
similarly to \r{sd8}
 
\section{Concluding Remarks and Outlook}
We have considered generalised supersymmetry algebras, 
including elements
transforming according to general Lorentz representations 
$(a,\dot a)$ 
possibly having spin greater
than one.

As an application of our analysis, we have generalised the notion of
\sdy, the paradigm of solvability for systems with four
independent variables, 
to the hyperspaces $\M$. We have thus obtained 
a hierarchy of solvable systems in 
an arbitrarily large number of variables. 

There are several directions in which our considerations  
afford generalisation. 

{\bf a)} In the complex setting, having in mind physical considerations, 
we have worked with
representations $(a,\dot a)$ of the algebra $su(2)\oplus
su(2)\simeq sp(1)\oplus sp(1)$ of the group $SO(4)$. 
As a possible generalisation,
this basic algebra could be extended to the direct sum of two algebras 
$g\oplus h$, for example, $sp(1)\oplus sp(n)$. Generalisations
involving algebras which are not direct sums may also be considered.

{\bf b)} We have worked with supercommuting coordinates satisfying 
\r{comYY}. Our formalism suggests generalisations
to either non-commutative geometry and/or to quantum type (deformed)
commutation relations, i.e. appropriate deformations
of the superalgebra of the $X$'s
and $Y$'s and of the gauge group.    

{\bf c)} We could extend the superalgebra of $X$'s and $Y$'s
to associative algebras containing suitable non-linear terms
in the right-hand sides of the supercommutators. However,
non-linear terms in the algebra $\A\ $ yield hyperspaces on
which the minimal coupling of gauge fields is not well defined.  

{\bf d)} We have restricted ourselves to flat spaces. A
generalisation to curved spaces using
{\it diffeomorphism-covariant} derivatives may also be considered,
with the $X$'s in $\A\ $ interpreted as vector fields 
spanning the tangent space.

\vskip 15pt\noindent
{\bf{Acknowledgments}}

One of the authors (J.N.) would like to thank Professors
S.~Randjbar-Daemi and F.~Hussain and the ICTP for hospitality while
this work was begun. 
The other (C.D.) thanks E.S.~Fradkin and M.A.~Vasiliev for encouraging
discussions. Both authors thank the Fonds National de la
Recherche Scientifique (Belgium) for partial support. 

\appendix
\section{Multiplicities}\la{mult}
\setcounter{equation}{0}\renewcommand\theequation{A\arabic{equation}}
\subsection{}
In this appendix we give the main formulas of the text modified
in such a way as to allow for multiplicities of the different
operators corresponding to the coordinates and 
the vector fields.
Let $N_{a,\dot a}^Z$
the multiplicity of the operators  $Z=Y$ or $Z=X$
of given behaviour
$a,\dot a$ under the Lorentz algebra. The set of coordinates 
$Y_{[A]}^{[\dot A]}(m)$ (resp. the set of
differential vector operators $X^{[\dot A]}_{[A]}(m)$)  
is indexed by an integer $m$ with $1\leq m \leq N_{a,\dot a}^Y$
(resp. $1\leq m \leq N_{a,\dot a}^X$). 
We expect that in the interesting cases $N_{a,\dot a}^X =
N_{a,\dot a}^Y$ i.e. that the number of coordinates is exactly
equal to the number of generalised derivatives (except perhaps
for the Lorentz generators themselves which don't necessarily
need their coordinate counterparts).

Using this notation, (\ref{comYY}) becomes
\be
\left [Y_{[A]}^{[\dot A]}(m),Y_{[B]}^{[\dot B]}(n)\right ]=0 
\label{comYYm}
\ee
and (\ref{comXX}) becomes
\bea
&& \left[ X_{[A]}^{[\dot{A}]}(m),X_{[B]}^{[\dot{B}]}(n)\right ]
\nonumber \\
    &=&\sum_{s=0}^{\min (2a,2b)}\ 
       \sum_{\dot s=0}^{\min (2\dot a,2\dot b)} \ 
       \sum_{p=1}^{\left(N_{a+b-s,\dot a+\dot b-\dot s}^X\right)}
  t(a,\dot a,m\,;b,\dot b,n\,;a+b-s,\dot a+\dot b-\dot s,p)
                              \nonumber   \\   
     &&\hskip 5.7 true cm S[A]S[\dot A]S[B]S[\dot B]
     \epsilon_{[\a_s\b_s]}\epsilon^{[\da_{\dot s}\db_{\dot s}]}
     X_{[C(s)]}^{[\dot C(\dot s)]}(p)
\label{comXXm}
\eea
where the structure constants $t$ not only have the 
Lorentz labels of the three tensors involved but also on their 
multiplicity indices.
Finally, (\ref{comXY}) becomes
\bea
 && \left[X_{[A]}^{[\dot A]}(m),Y_{[B]}^{[\dot B]}(n)\right ]
\nonumber \\
  &=&\sum_{s=0}^{\min (2a,2b)}\sum_{\dot s=0}^{\min (2\dot a,2\dot b)}
    \sum_{p=1}^{\left(N_{a+b-s,\dot a+\dot b-\dot s}^Y\right)}
     u(a,\dot a,m\,;b,\dot b,n\,;a+b-s,\dot a+\dot b-\dot s,p)
                             \nonumber \\
     &&\hskip 5.3 true cm S[A]S[\dot A]S[B]S[\dot B]
        \epsilon_{[\a_s\b_s]}\epsilon^{[\da_{\dot s}\db_{\dot s}]}
     Y_{[C(s)]}^{[\dot C(\dot s)]}(p)    
                            \nonumber \\ 
      &&+\  c(a,\dot a\,;m,n)\delta^{ab}\delta_{\dot a\dot b}
      S[A]S[\dot A]
      \epsilon_{[\a_{2a}\b_{2a}]}
      \epsilon^{[\da_{2\dot a}\db_{2\dot a}]}
\label{comXYm}
\eea
where the structure constants $u$ 
depend also on the multiplicity indices involved.
For a given $a$ and a given $\dot a$, the central term
parameters $c(a,\dot a\,;m,n)$ form a 
$N_{a,\dot a}^X\times N_{a,\dot a}^Y$ matrix. 
We expect that
in the interesting cases this
matrix is square and non singular and that it can be brought to the unit
matrix $\delta_{mn}$ by redefining suitable linear combinations of the
$X$'s and of the $Y$'s as the basic operators. 

As far as the Jacobi identities are concerned, the
multiplicity indices have to be taken into account. In
particular, a summation
on the `internal' multiplicity index has to be included in the
generalisation of, for instance, \r{jacobitt}. 

\subsection{}

When a certain Lorentz representation $(a,\dot a)$ occurs multiply, the
corresponding tensors can be linearly combined.
Using the matrices $A\in GL(N^X_{a\dot a},\C)$ 
for the $X$'s and  $B\in GL(N^Y_{a\dot a},\C)$ for the $Y$'s,
the allowed transformations are
\be
X'(m')=A(m',m)X(m),\quad Y'(n')=B(n',n)Y(n)
\ee 
where we have indicated only the multiplicity index. 
(Note that, in the real
setting, the field $\R$ should be used rather than $\C$).

We may use this freedom, for instance, to put the, a priori complex,
matrix $c(a,\dot a\,;m,n)$, which transforms as
\be
c'(m',n')=A(m',m)B(n',n)c(m,n)\quad 
    \Leftrightarrow \quad c'=AcB^t,
\ee
in the canonical form
\be               
  \wt c(m,n) =   \pmatrix{ 1_{(r,r)} & 0_{(r,s)}\cr
                       0_{(t,r)} & 0_{(t,s)}\cr} 
\la{cf}
\ee
where $1_{(r,r)}$ is the unit $r\times r$ matrix and $0_{(p,q)}$ 
is the zero matrix with $p$ rows and $q$ columns. The number of $X$'s,
i.e. $(r+t)$, can be different from the number of $Y$'s, i.e. $(r+s)$. 
The stability group of this canonical $\wt c(m,n)$ 
has matrices of the form
\be 
A=\pmatrix{        A^1_{(r,r)} & A^2_{(r,t)}  \cr
                   0_{(t,r)} & A^4_{(t,t)}  \cr}
,\quad 
B=\pmatrix{       B^1_{(r,r)} & B^2_{(r,s)}  \cr
                  0_{(s,r)} & B^4_{(s,s)}  \cr}
,\quad
B^{1t}={(A^1)}^{-1}
\ee
where $A^1$, $A^4$ and $B^4$ are arbitrary invertible matrices.

This stability group
can be used to put the $t$'s
and/or the $u$'s which transform as 
(only indicating the multiplicity index)
\be
\arr          
        t'(m',n',p') &=& A(m',m)\ A(n'n)\ A^{-1}(p',p)\ t(m,n,p) \\[5pt]
        u'(m',n',p') &=& A(m',m)\ B(n'n)\ B^{-1}(p',p)\ u(m,n,p)   
\ea 
\ee
into some canonical form.

We could also consider non-linear transformations among
the $X$'s or the $Y$'s which preserve the Lorentz transformation
properties. However we feel that such transformations have little
physical significance.

\section{The $SU(2)\otimes SU(2)$ Harmonics}
\la{harmapp}
\setcounter{equation}{0}\renewcommand\theequation{B\arabic{equation}}
\def\op{\oplus} 
\def\om{\ominus}
\def\dpp{D^{++} }
\def\dopp{D^{\op\op} }
\def\vpp{V^{++} }
\def\vopp{V^{\op\op} }
\def\npp{\n^{++} }
\def\nopp{\n^{\op\op} }
\def\dmm{D^{--} }
\def\domm{D^{\om\om} }
\def\vmm{V^{--} }
\def\vomm{V^{\om\om} }
\def\nmm{\n^{--} }
\def\nomm{\n^{\om\om} }

In this appendix we outline the harmonic space notation \c{harm}.
We use two sets of harmonics parametrising auxiliary spaces. 
These are commuting 
spinors $u^+_\da , u^-_\da$ and $v^\op_\a , v^\om_\a$, 
which satisfy the constraints
\be
u^{+\da} u_\da^{-} = 1  ~,~ v^{\oplus\a} v_\a^{\ominus} = 1\ .
\ee

In the euclidian case, an $SU(2)$ matrix can be written
\be
U=\left({\matrix {\a  & -\b^* \cr
                  \b   &   \a^*    }}   \right),\ \   {\rm{with}}
\quad \mid\a\mid^2+\mid\b\mid^2=1\ . 
\ee
Let 
\be
V=\left( {\matrix {e^{i\phi}     & 0     \cr
                      0         & e^{-i\phi}}}
                 \right)
\ee
be a $U(1)$ subgroup of $SU(2)$. The quotient $SU(2)/U(1)$ given
by the equivalence classes is a 2-sphere and can be described by
the spinors
\be
u^{+\da}= \left({\matrix {\a   \cr
                            \b     }}   \right),\ \  {\rm{and}}
\quad u^{-}_{\da}={u^{+\da}}^*\Leftrightarrow
u^{-\da}=\left({\matrix { -\b^* \cr
                             \a^*   }}   \right)
\ee
up to their respective phases $u^{+\da}\equiv \exp(i\phi)u^{+\da}$
and $u^{-\da}\equiv \exp(-i\phi)u^{-\da}$.

Vector fields on these auxiliary spaces are given by
\be
\arr  D^{++} =\ u^{+ \da} \der {u^{- \da}}, &\quad &
         D^{\op\op}= v^{\op \a} \der {v^{\om \a}} \\[5pt]
         D^{--} =\ u^{- \da} \der {u^{+ \da}}, &\quad &
         D^{\om\om}= v^{\om \a} \der {v^{\op \a}} \\[5pt]
 D^{+-} = u^{+\da} \der {u^{+\da}} - u^{-\da} \der {u^{-\da}}, &\quad &
 D^{\op\om} = v^{\op\a} \der {v^{\op\a}} - v^{\om\a} \der {v^{\om\a}} 
\la{flat}
\ea
\ee
and they satisfy two commuting SU(2) algebras.


\goodbreak
\end{document}